\documentstyle[11pt]{article}
\input epsf.tex
\normalbaselines
\begin{document}
\bibliographystyle{unsrt}

\def\bea*{\begin{eqnarray*}}
\def\eea*{\end{eqnarray*}}
\def\ba{\begin{array}}
\def\ea{\end{array}}
\count1=1
\def\be{\ifnum \count1=0 $$ \else \begin{equation}\fi}
\def\ee{\ifnum\count1=0 $$ \else \end{equation}\fi}
\def\ele(#1){\ifnum\count1=0 \eqno({\bf #1}) $$ \else \label{#1}\end{equation}\fi}
\def\req(#1){\ifnum\count1=0 {\bf #1}\else \ref{#1}\fi}
\def\bea(#1){\ifnum \count1=0   $$ \begin{array}{#1}
\else \begin{equation} \begin{array}{#1} \fi}
\def\eea{\ifnum \count1=0 \end{array} $$
\else  \end{array}\end{equation}\fi}
\def\elea(#1){\ifnum \count1=0 \end{array}\label{#1}\eqno({\bf #1}) $$
\else\end{array}\label{#1}\end{equation}\fi}
\def\cit(#1){
\ifnum\count1=0 {\bf #1} \cite{#1} \else 
\cite{#1}\fi}
\def\bibit(#1){\ifnum\count1=0 \bibitem{#1} [#1    ] \else \bibitem{#1}\fi}
\def\ds{\displaystyle}
\def\hb{\hfill\break}
\def\comment#1{\hb {***** {\em #1} *****}\hb }

\newcommand{\ZZ}{\hbox{Z\hspace{-3pt}Z}}
\newcommand{\NZ}{\hbox{I\hspace{-2pt}N}}
\newcommand{\RZ}{\hbox{I\hspace{-2pt}R}}
\newcommand{\BZ}{\,\hbox{I\hspace{-2pt}B}}
\newcommand{\CZ}{\,\hbox{I\hspace{-6pt}C}}
\newcommand{\PZ}{\hbox{I\hspace{-2pt}P}}
\newcommand{\HZ}{\hbox{I\hspace{-2pt}H}}
\newcommand{\EZ}{\hbox{I\hspace{-2pt}E}}

\vbox{\vspace{38mm}}
\begin{center}
{\LARGE \bf Gauge Symmetry and Integrable Models}\\[5mm]
Shao-shiung Lin \\
{\it Department of Mathematics,\\ 
Taiwan University \\ Taipei, Taiwan \\
 (e-mail: lin@math.ntu.edu.tw)} \\ [3 mm]
Oktay K. Pashaev \\
{\it Joint Institute for Nuclear Research \\ 
Dubna 141980, Russian Federation .\\
 (e-mail: pashaev@vxjinr.jinr.dubna.ru )} \\ [3 mm]
Shi-shyr Roan\footnote{Supported in part by 
the NSC grant of Taiwan.}\\{\it Institute of Mathematics \\ Academia Sinica \\ 
Taipei , Taiwan \\ (e-mail: maroan@ccvax.sinica.edu.tw)} \\[5mm]
\end{center}

\begin{abstract} 
We establish the isomorphism between a nonlinear $\sigma$-model and the 
abelian gauge theory on an arbitrary curved background, which allows us to derive 
integrable models and the corresponding Lax representations from gauge 
theoretical point of view. In our approach the spectral parameter is related 
to the global degree of freedom associated with the conformal or 
Galileo transformations of the spacetime. The B$\ddot{\rm a}$cklund 
transformations are derived from Chern-Simons theory where the 
spectral parameter is defined in terms of the extract compactified space 
dimension coordinate.

\par \vspace{.2in} \noindent
1996 PACS: 02, 03, 11  
\end{abstract}

\vfill
\eject

\section{Introduction}

The aim of this paper is to present in a clear form the systematic 
method outlined in Ref. \cite{P96} of generating Lax-form and the 
spectral parameter of  
integrable models through the gauge theory. For the gauge group 
$SU(2)$, the method leads both the non-relativistic and relativistic 
integrable equations in 1-space and 1-time dimension, whose fundamental 
examples are respectively the celebrated nonlinear 
Schr$\ddot{\rm o}$dinger (NLS) 
equation and the Liouville equation. 
In the theory of classical integrable systems, the zero-curvature 
formulation has been a crucial step to elucidate the hidden symmetry of 
the equation \cite{Po} \cite{EP} \cite{O}. A large class of solvable 
systems originated in the gauge theory. 
Of course such systems represent only a restricted, 
but nevertheless rich, class of integrable equations of 
physical interests.
It has been well-known that the relation of gauge equivalence   
exists between 
various models like the continuum Heisenberg spin chain and 
NLS \cite{FT}. Historically, these like relations were 
established on the level of Lax or zero-curvature representations  
\cite{ZS} \cite{ZT} \cite{PS}, 
but without admitting an extention to the nontrivial backgroud of higher 
dimensions. From another site the $\CZ\PZ^n$  
$\sigma$-model representation \cite{ALV} allows one to associate the 
model with a gauge field theory using the auxiliary fields. In this 
paper we shall focus on some well-known models of which 
the Lax forms already exist 
in literature, and establish the
equivalent relations among them through the  
$SU(2)$ or $SU(1,1)$-gauge theory. 
As it is known for integrable models, the  Lax representation 
is the fundamental procedure for the solvablity of the equation. However 
the achievement of Lax forms with a 
spectral parameter usually relies on the experience of 
expertise. It is desirable to have some additional mathematical or 
physical grounds for the understanding of the qualitative nature of 
existing results. In the gauge theory approach of  
integrable systems, we are able to associate  
a spectral parameter with the global degree of freedom 
of the equations connected to some special automorphisms of  
spacetime, e.g. the Galileo transformations for NLS. Furthermore 
the B$\ddot{\rm a}$cklund 
transformations can be derived from the higher dimensional 
Chern-Simons theory with the interpretation of the extract space 
dimension coordinate as the 
spectral parameter. 

Our paper is organized as follows: In Section 2 we 
review the general  
properties of connections and homogenous  
spaces, which will be relevent to the 
discussion of this paper. In what follows, we present a systematic 
approach to integrable models related to the gauge 
theory of $SU(2)$ or $SU(1,1)$. 
In Section 3 we establish the general correspodence between $S^2$ 
$\sigma$-models and abelian gauge theories with background fields 
derived from the $SU(2)$-flat connections. Models with certain  
special constraints will be treated in the next three sections. 
In Section 4  we discuss the 
self-dual equation of a Riemann surface and its corresponding 
abelian gauge system. In particular for the 
complex plane, it gives rise the solutions of Liouville equation. 
In Section 5 the (conformal) sinh-Gordon 
equation and 
and its various equivalent forms will be derived. 
We shall relate the spectral parameter of the sinh-Gordon 
equation with conformal symmetries of the equation.  
In Section 6 we give a gauge theory description of NLS equation, 
and establish its 
equivalent relation with the Heisenberg model as the 
corresponding $\sigma$-model. The Zakharov-Shabat representation of NLS
equation will be derived 
from the Galileo invariance of the equation. In Section 7 we 
present an interpretation of the B\"{a}cklund transformation of 
NLS through the three dimensional Chern-Simons theory by the technique 
of dimensional reduction, a treatment in the framework of       
BF-type gauge theory in \cite{P} \cite{P96}. We 
reinterpret the   
spectral parameter of NLS equation as 
the extract compactified space dimension coordinate. 
In Section 8 we present the correspodence bewteen hyperbolic 
non-compact $\sigma$-models and the abelian gauge models  
derived from $SU(1,1)$-flat connections, in particular   
the Liouville equation, and its  relation 
with the cylindrical symmetry solutions of the 4-dimensional self-dual 
Yang-Mills equation obtained by E. Witten in \cite{W}. 
In the conclusion we discuss some physical ideas to explain our results.

\par \vspace{.2in} \noindent
{\bf Notations.}
\par  \noindent
We shall use the following notations unless we mention specifically.
Let $X$ be a (differentiable) manifold, and ${\bf L}$ be 
a complex Hermitian line bundle over $X$. We use $x^\mu$ as the local 
coordinates of $X$.  By a local open subset of 
$X$, we shall always mean a contractible open set $U$ endowed with a coordinate 
system. Denote:
 
$\Omega^k(X)$ = the vector space of real  $k$-forms on $X$. 
  
$\Omega^k(X)_{\CZ} = \Omega^k(X) \otimes_{\RZ} \CZ$.  

$\Omega^k(X, {\bf L})$= the vector space of ${\bf L}$-valued 
$k$-forms on $X$. \par \noindent 
For $s \in \Omega^k(X, {\bf L})$, we shall 
denote the complex conjugate of $s$ by 
$s^{\dag} \in \Omega^k(X, {\bf L}^*)$. 
For manifolds $X, Y$, with base elements $x_0 \in X, y_0 \in Y$, we 
denote 

${\cal C}^{\infty} (( X, x_0 ) , (Y , y_0)  ) $ = the set of all 
${\cal C}^{\infty}$-maps $\varphi$ from $X$ to $Y$ with 
$\varphi(x_0) = y_0$. For simplicity, we shall call 
$(X, x_0)$ a marked manifold $X$ (with the base element $x_0$), and 
 $\varphi$ a marked differentiable map from $X$ to $Y$. In 
what follows, we systematically  
write ${\cal C}^{\infty} (X, Y)$ instead of 
${\cal C}^{\infty} (( X, x_0 ) , (Y , y_0)  )$  
if no confusion could arise. 
For a Lie group the base element will always be the identity element.
   
${\rm Diff}(X)$ ( = ${\rm Diff}(X, x_0)$  ) = the group of 
automorphisms of $X$ preserving the base element $x_0$.
 \par \noindent

\section{Preliminaries : Connection and Curvature}
A systematic review of connection and curvature cannot be presented in a 
regular paper. Here we simply recall some basic concepts and a few definitions, 
in order to fix the notations used in this work. We follow in this respect 
Ref. \cite{Chern} \cite{KN}.  
Let $G$ be a  semi-simple Lie group with the Lie algebra ${\bf g}$, and  
${\rm Aut}({\bf g})$ be the identity component of the group of Lie-automorphisms 
of ${\bf g}$. 
The negative Killing form defines 
an inner product structure $< *, * >$ of  ${\bf g}$, invariant under 
the adjoint action of $G$. Let ${\rm Der}({\bf g})$ be the algebra of derivations of 
${\bf g}$. It is known that  
\[
{\bf g} \simeq {\rm ad}({\bf g}) = {\rm Der}({\bf g}) \ , \ \ 
\]
and one has the exact sequence:
\bea(l)
1 \longrightarrow {\rm Z}(G) \longrightarrow G 
\stackrel{\rm Ad}{\longrightarrow} {\rm Aut}({\bf g}) \longrightarrow 1 
\elea(GAd)
where ${\rm Z}(G)$ is the center of $G$.
Then we have    
\[
\begin{array}{l}
{\rm Aut}({\bf g})  \subset SO ( {\bf g}) := \{ \rho \in SL({\bf g}) \ | \ \ 
< \rho ( x ) , \rho ( y ) > = 
< x , y > , \ \forall x , y \in {\bf g}  \} \ , \ \\ 
{\rm Der}({\bf g}) \subset so ( {\bf g} ) = {\rm Lie \ algebra \ of \ } 
SO({\bf g}) \ ,
\end{array}
\]
and we shall always consider ${\rm Aut}({\bf g})$ as a closed subgroup of 
$SO ( {\bf g}) $ in what follows.
Let $ ( e^1, \ldots, e^n )$ 
be an orthonormal basis of ${\bf g}$ with respect to $<*,*>$. The Lie algebra ${\bf g}$ 
is defined by
$$ 
[ e^i, e^j ] = e^k c^{ij}_k  \ , \ \ \  1 \leq i, j, k \leq n \ ,
$$
where the symbol $c^{ij}_k$ is a 
structure constant of ${\bf g}$ and is anti-symmetric in its indices. 
Let $(e_1, \ldots, e_n )$ be the basis of ${\bf g}^*$ dual to $(e^1, \ldots , 
e^n)$. As ${\bf g}$ is isomorphic to ${\bf g}^*$ via $<*,*>$,
equivariant  with 
respect to $G$-adjoint actions, we shall 
make the identification:  ${\bf g} = {\bf g}^*, 
e^j = e_j$. 
Using the basis $e^j$, one regards Aut$({\bf g})$ as 
a subgroup of $SO(n_+, n_-)$ by the following relation, 
where $(n_+, n_-)$ is the signature of $<*,*>$,
\bea(lll)
 {\rm Ad}(g) \leftrightarrow &
R(g)= ( R(g)_j^k) , & {\rm with } \ \ 
{\rm Ad}(g)( e^1, \ldots ,  e^n ) = 
( e^1, \ldots , e^n ) R(g) \ , \ \ g \in  G \ \ . 
\elea(subg)
Now it is easy to see the following lemma.
\par \vspace{0.2in} \noindent
{\bf Lemma 1. } For a linear map $l$ of ${\bf g}$, 
\[
l : ( e^1, \ldots, e^n )  \mapsto ( e^1, \ldots, e^n )  M , \ \ M = 
( M_i^j ) \ , 
\]
the following conditions are equivalent:

(I) $l  \in {\rm Der}({\bf g})$ .

(II) $
M_k^sc_s^{ij} = M^i_sc^{sj}_k - M^j_sc^{si}_k $ for all $i , j , k$.

(II) There exist real numbers, $\alpha_1, \ldots, \alpha_n$, such that 
$M^i_k = c^{ji}_k \alpha_j$ for all $i, k$.
$\Box$ \par \vspace{0.2in} \noindent 
Let $P$ be a principal $G$-bundle over a manifold $X$,
$$
G \longrightarrow P \longrightarrow X \ .
$$
The bundle $P$ induces a principal 
${\rm Aut}({\bf g})$-bundle $P'$ and a vector bundle 
${\rm ad}(P) $ over $X$ associated to the 
adjoint representation of $G$ on ${\bf g}$:  
$$
P' = P \times_G {\rm Aut}({\bf g}) \ , \ \ 
{\rm ad}(P) = P \times_G {\bf g}  \ . 
$$ 
Denote $\Omega^*(X, {\rm ad}(P))$ the set of differential forms with 
values in ${\rm ad}(P)$. 
The fibers of  
${\rm ad}(P)$ are endowed with a natural Lie algebra structure 
from the Lie algebra ${\bf g}$. The ${\rm Aut}({\bf g})$-bundle $P'$ can be regarded as the 
bundle of frames $(v^1, \ldots , v^n)$ of ${\rm ad }(P)$ with the fixed 
Lie structure constants. 
Let $J$ be a connection on $P$ over a manifold $X$.   
It gives rise to a 
${\rm Aut}({\bf g})$-conncetion $J_{\rm ad}$ on the bundle $P'$, 
and a covariant 
differental $D_{\rm ad}$ on $\Omega^*(X, {\rm ad}(P))$, 
$$
\begin{array}{l}
D_{\rm ad} ( \phi ) =  d(\phi) + [ J , \phi ]  \ , \ \ \phi \in 
\Omega^*(X, {\rm ad}(P)) \ . 
\end{array}
$$
For a local open subset $U$ of $X$, the local 
expressions of 
$J, J_{\rm ad}$ and $D_{\rm ad}$ are given by
\bea(lll)
J = &   e^j \omega_j  = J_{\mu} dx^\mu 
\in {\bf g} \otimes \Omega^1(U) , &
\omega_j =  J_{j,\mu} dx^{\mu} \in \Omega^1(U) \ , \ 
J_\mu = e^j J_{j,\mu} \in \Omega^0(U) \otimes {\bf g} \ , \\
J_{\rm ad} = & e^k e_{i} (\omega_{\rm ad})^i_k 
\in {\rm Der}({\bf g}) \otimes \Omega^1(U) \ ,& (\omega_{\rm ad})^i_k =  
 (J_{\rm ad})_{k, \mu}^idx^{\mu} \ , \ 
(J_{\rm ad})_{k, \mu}^i : =  c^{ji}_k  J_{j,\mu} \ ,
\elea(conn)
and 
$$
\begin{array}{l}
D_{\rm ad} ( e^if_i ) = e^i d(f_i) + (D_{\rm ad} e^i) f_i \ , 
\ \ \ \ f_j \in \Omega^*(X) \ , \\
D_{\rm ad} e^i  = [e^j \omega_j ,  e^i ]  =  
e^k (J_{\rm ad})_{k, \mu}^i dx^\mu \in {\bf g} \otimes \Omega^1(X)   .
\end{array}
$$
The curvature $F(A)$ is a 2-form with values in ${\rm ad}(P)$:
$$
F(J) = d J + \frac{1}{2}[ J , J ]  \in \Omega^2(X, {\rm ad}(P)) \ .
$$
Recall that two connections 
$$
J = e^j\omega_j \ , \ \ \tilde{J} = e^j \tilde{\omega}_j \ ,
$$
are gauge equivalent if there exists an element
$h \in {\cal C}^\infty (X,  G )$ such that
$$
\tilde{J}  = h^{-1} dh + 
{\rm Ad}(h)^{-1}(e^j)\omega_j \ , \ \ 
\tilde{\omega}_j =  \tilde{J}_{j,\mu} dx^{\mu} \ , \ \ 
\tilde{J}_{j,\mu} = J(h)_{j, \mu} + R(h^{-1})_j^k J_{k, \mu} \ . 
$$
This implies the corresponding Aut$({\bf g})$-connections, 
$(\tilde{J}_{\rm ad})_{k, \mu}^i d x^\mu$, 
$(J_{\rm ad})_{k, \mu}^i d x^\mu$, are also equivalent.
It is known that the diffeomorphism group ${\rm Diff}(X)$ acts on the 
space of $G$-connectons 
in a canonical manner. In fact, for $\varphi \in {\rm Diff}(X)$ and a connection 
$J$ on a principal $G$-bundle, $\varphi^*J$ is a conncection on the 
pull-back bundle $\varphi^*P$.   
For a function $g \in {\cal C}^\infty ( U , G)$, 
it defines a current  by 
$$
 J(g)_{ \mu}dx^{\mu} := g^{-1} dg  \in {\bf g} \otimes \Lambda^1 \ , \ \ 
J(g)_{\mu} = e^j J(g)_{j, \mu} \ .
$$ 
A connection $J$ is flat if $F(J) = 0$. Its  
local expression is given by 
$$
\begin{array}{lll}
&J =  e^j \omega_j = g^{-1} dg \ , \ \ & {\rm for } \ \
 g \in {\cal C}^\infty ( U , G) \ , \ 
\ {\rm i.e.} \ J_{\mu} = J(g)_{\mu} \ ,  \\
 \Longleftrightarrow & 
d\omega_i + \frac{1}{2} c^{jk}_i \omega_j \wedge \omega_k = 0 
 \ , & {\rm i.e. } \  \ 
\partial_{\mu} J_{\nu} - \partial_{ \nu} J_{\mu} + 
[ J_{\mu} ,  J_{\nu} ] = 0 \ .
\end{array}
$$
For two gauge equivalent connections, $J, \tilde{J}$ , 
the flat condition of one implies the other. In fact, one has 
$$
J = g^{-1}dg \ \ 
\Longleftrightarrow \ \ \tilde{J} = (gh)^{-1} d (gh ) \ \ .
$$ 
As the $G$-bundle $P$ is a finite cover over the 
${\rm Aut}({\bf g})$-bundle $P'$ with the 
covering group $Z(G)$, 
$$
\begin{array}{lcc}
 P & \longrightarrow & P/{\rm Z}(G) = P' \\ 
\downarrow & & \downarrow  \\
X & = & X \ \ \ ,
\end{array}
$$
the flat conditions of connections, $J$ , $J_{\rm ad}$, 
are equivalent. 
For a simply-connected manifold $X$ with a base element $x_0$, 
there is an    
one-to-one correspondence of the following data, 
 equivariant under the action of ${\rm Diff}(X)$:

(I) A flat $G$-connection on $P$.

(II) A flat Aut$({\bf g})$-connection on $P'$.

(III) $g \in {\cal C}^\infty ( ( X ,  x_0) , (G, 1) )$.

(IV) $m \in {\cal C}^\infty ( ( X ,  x_0) , ({\rm Aut}({\bf g}), 1) )$.
\par \noindent 
Here an element $\varphi \in {\rm Diff}(X)$ acts on $g , m$ by
$$
\begin{array}{ll}
( \varphi , g ) \mapsto 
\varphi^*g \ ,& 
\varphi^*g(x) := (g(\varphi(x_0))^{-1}g(\varphi(x)) \ , \ 
{\rm for }  \ x \in X  \ , \\ 
( \varphi , m ) \mapsto 
\varphi^*m \ , & 
\varphi^*m(x) := (m(\varphi(x_0))^{-1}m(\varphi(x)) \ , \ 
{\rm for } \ x \in X  \ .
\end{array}
$$
The functions $ g , m $, in (III), (IV), are the gauge functions of  
flat connections of (I), (II), respectively. 
Note that under this situation, the bundle 
$P$ is necessarily a trivial one.   
With $ R(g) =  ( m_k^i) $ in (\req(subg)), we have the following result.
\par \vspace{0.2in} \noindent
{\bf Proposition 1.} Let $X$ be a 
simply-connected marked manifold ( with a base 
element $x_0$), 
and $J_{\mu} = e^j J_{j, \mu} 
\in \Omega^0(X) \otimes {\bf g} , 1 \leq j \leq n $. Then the following conditions are 
equivalent:

(I) 
\bea(l)
 \partial_{\mu} J_{ \nu} - \partial_{ \nu} J_{ \mu} + [  
J_{ \mu},  J_{ \nu} ] = 0  \ \ , \ \ {\rm i.e.} \ \ 
\ [ \partial_\mu + J_\mu , \partial_\nu + J_\nu ] = 0 \ , \ \ {\rm \ 
for \ all } \ \mu, \nu.
\elea(0F)

(II) There exists an element $m = e^ke_i m_k^i \in 
{\cal C}^\infty (X, {\rm Aut}({\bf g}))$ 
( := ${\cal C}^\infty (( X ,  x_0) , ({\rm Aut}({\bf g}), 1))$ )   
such that 
$$
(\partial_{\mu} m_k^i) = (m_k^i) ( (J_{\rm ad})_{k, \mu}^i  ) \ , \ \ \  
(J_{\rm ad})_{k, \mu}^i = c_k^{ji}J_{j,\mu} \ ,
$$
i.e.  
\bea(l)
\partial_{\mu}( m^1, \ldots , m^n ) = 
( m^1, \ldots , m^n ) ( (J_{\rm ad})_{k, \mu}^i  ) \ ,
\elea(Adeq)
where $m^j$ is the $j$-th column vector of the matrix 
$(m_k^i)$. Furthermore, the above correspondence is 
equivariant under actions of ${\rm Diff}(X)$.
$\Box$ \par \vspace{.2in} \noindent
For the group $G = U(1)$, we shall identified the Lie algebra of 
$U(1)$ with $i\RZ$ through the usual exponential map,
$$
0 \longrightarrow  2 \pi i \ZZ 
\longrightarrow i \RZ \stackrel{\rm exp}{\longrightarrow} U(1) \longrightarrow 1 \ .
$$  
For a connection $A$ on a principal $U(1)$-bundle $P$ over a 
manifold $X$, ${\rm ad}(P)$ is the trivial bundle with 
$D_{\rm ad}$= the ordinary  differential $d$. 
However, through the canonical 
emdedding $U(1) \subset \CZ^* = GL_1(\CZ)$, there is a Hermitian 
(complex) line bundle ${\bf L}$ over $X$ associated to $P$ with a 
covariant differential $d_A$ on sections of ${\bf L}$ and its extension   
on ${\bf L}$-valued forms,
$$
d_A : \Omega^*(X, {\bf L}) \longrightarrow \Omega^*(X, {\bf L}) \ .
$$
The Chern class $c_1({\bf L})$ 
of ${\bf L}$ is the integral class represented by the curvature $
\frac{i}{2\pi}F(A)$:
$$
c_1({\bf L}) = \frac{i}{2\pi} F(A) \ \in \ {\rm H}^2(X, \ZZ) \ .
$$
On a local open set $U$,   
$A$ is expressed by an 1-form $i V$ for $ V \in \Omega^1(U)$. In this situation, 
we have $F(A) = dA$, and the covariant differential $d_A$, 
$$
d_A : = d + A : \Omega^*(U)_{\CZ} 
\longrightarrow \Omega^*(U)_{\CZ} \ .
$$ 
In a local coordinate 
system $(x^1, \ldots , x^n )$, we shall write 
$$
d_A ( f ) = d x^\mu (\partial_A)_{\mu} f \ , \ \  
{\rm for } \ f \in \Omega^0(U)_{\CZ} \ .  
$$

Let $H$ be a closed subgroup of $G$,  $S$ be the $H$-orbit space $G/H$ 
with the base element $e$ equal to the identity coset, and  
$\pi: G \longrightarrow S$ the natural projection. 
 We have the  
exact sequence of sets of marked maps, 
equivariant under the actions of  Diff$(X)$,
$$
0 \longrightarrow {\cal C}^\infty(X, H) 
\longrightarrow {\cal C}^\infty(X, G) 
\stackrel{\pi_*}{\longrightarrow} {\cal C}^\infty(X, S) 
\stackrel{\delta}{\longrightarrow} {\rm H}^1( X, \Omega^0(H) ) 
\longrightarrow \cdots \ .
$$  
For $s \in {\cal C}^\infty (X, S)$ with $\delta(s) = 0$, we have 
 $s = \pi \cdot g $ for a function $g \in {\cal C}^\infty(X, G)$, 
 unique up to the multiplication of an element $h $ in ${\cal C}^\infty(X, H)$, 
i.e. $g \sim gh$. One can assign the element $s$ a 
flat $G$-connection 
$e^jJ_{j, \mu} dx^\mu$ corresponding to $g$, then modulo the qauge 
equivalent relation induced by elements of ${\cal C}^\infty(X, H)$.
Hence we have the following result: 
\par \vspace{0.2in} \noindent
{\bf Proposition 2.} For a simply-connected marked manifold $X$, there 
is an one-to-one correspondence of the 
following data, which are equivariant under the actions of 
Diff$(X)$:

(I) $s \in {\cal C}^\infty (X, S)$ with $\delta(s) = 0$.

(II) $ J_{\mu} = e^j J_{j, \mu} dx^\mu  
\in {\bf g} \otimes \Omega^0(X)  $, 
a flat connection modulo the relation, $J_{\mu} \sim 
\tilde{J}_\mu$,  
$$
\tilde{J}_{j,\mu} = J(h)_{j, \mu} + R(h^{-1})_j^k J_{k, \mu} \ , \
\ {\rm for  \ } h \in {\cal C}^\infty(X, H) \ .
$$
The relation of $s$, $ J_{\mu} $, is given by 
$$
s = \pi_* ( g ) \ , \ \ g^{-1} dg = e^j J_{j,\mu} d x^\mu \ \ \ 
{\rm for \ some \  } g \in {\cal C}^\infty (X, G) \ .
$$
$\Box$ \par \vspace{.2in} \noindent
{\bf  Remark.} Under the 
situation where $H$ is abelian and ${\rm H}^2( X, \ZZ )=0$, 
one has ${\rm H}^1( X, \Omega^0(H) )=0$, then the condition 
(I) simply means $s \in {\cal C}^\infty (X, S)$.

\section{Flat SU(2)-Connection and S$^2 \ \sigma$-model}
In this section, we set the Lie algebra 
${\bf g} = su(2)$, where the negative Killing form is defined by 
$$
<x, y> = \frac{-1}{2} {\rm Tr}( {\rm ad } x \ {\rm ad } y   ) \ , \ \ 
x, y \in {\bf g} ,
$$
An orothonormal basis of ${\bf g}$ consists of the elements: 
$$
e^j = \frac{i}{2} \sigma^j \ , \ \ j=1,2,3 \ . 
$$
Here $\sigma^j$ are Pauli matrices,
\[ \sigma^1 = \left( \begin{array}{cc}
          0 & 1 \\
          1 & 0
         \end{array}   \right) , \ \
\sigma^2 = \left( \begin{array}{cc}
          0 & -i \\
          i & 0
         \end{array}   \right) , \ \
\sigma^3 = \left( \begin{array}{cc}
          1 & 0 \\
          0 & -1
         \end{array}   \right) . \ \ \]
Note that $\sigma^{\pm}, \frac{ \sigma^3}{2}$, form the standard basis of the Lie
algebra $sl_2(\CZ)$ with 
\[ \sigma^+ = \frac{\sigma^1 + i \sigma^2}{2} =
\left( \begin{array}{cc}
          0 & 1 \\
          0 & 0
         \end{array}   \right) , \ \ \ \
\sigma^- =  \frac{\sigma^1 - i \sigma^2}{2} =
\left( \begin{array}{cc}
          0 & 0 \\
          1 & 0
         \end{array}   \right) . \]
The Lie algebra ${\bf g}$ is given by 
$$
[ e^j , e^k ] =  - \epsilon_{jkl}e^l \ , 
$$
where $\epsilon_{jkl}$ are antisymmetric with $\epsilon_{123} = 1$. 
We have the equality,  Aut$({\bf g})$= $SO({\bf g})$ , which will be identified 
with $SO(3)$ via the basis $e^j$. The  sequence (\req(GAd)) 
becomes 
$$
1 \longrightarrow \{ {\pm 1} \} \longrightarrow SU(2) 
\longrightarrow SO(3) \longrightarrow 1 \ .
$$
Associated to an element of $SO(3)$, 
$(m^i_k) = (m^1, m^2, m^3)$,
we shall denote 
$$
\vec{S} = m^3 \ , 
$$
which is an element of the unit sphere $S^2$ in ${\bf g}$. We regard 
$(m^1, m^2)$ as a  positive orthonormal frame of the tangent space 
of $S^2$ at $\vec{S}$.
Note that $m^3 = m^1 \times m^2$. Hence we can  
identify $SO(3)$ with the unit tangent bundle of $ S^2$, 
i.e. 
$$
SO(3)  = \{ 
( \vec{S}, \vec{t} ) \in \RZ^3 \times \RZ^3 
\ | \ \vec{S}^2 = \vec{t}^2 = 1 \ , \ \vec{S}\cdot \vec{t} = 0 \ 
\} .
$$ 
For an element $( \vec{S}, \vec{t} ) \in SO(3)$, we shall  
denote 
$$
\vec{t}_+ = \vec{t} + i ( \vec{S}\times \vec{t} ) \ , \ 
\vec{t}_- = \vec{t} - i ( \vec{S} \times \vec{t} ) \ \ .
$$
The following relations hold: 
$$
\begin{array}{l}
\vec{t}_+ \cdot \vec{t}_+ = \vec{t}_{-} \cdot \vec{t}_- = 0 \ , \ \ \ \ 
\ \ \vec{t}_+ \cdot \vec{t}_{-} = 2 \ , \\
\vec{t}_+ \times  \vec{S} = i \vec{t}_+ \ , \ \ \   
\vec{t}_- \times  \vec{S} = - i \vec{t}_- \ ,  \ \ \ 
\vec{t}_- \times   \vec{t}_+ = 2i \vec{S} \ .
\end{array}
$$
The standard metric of $S^2$ together with the involution,   
$$
\vec{t} \mapsto \vec{S}\times \vec{t} \ , \ \ \ 
\vec{S}\times \vec{t} \mapsto -\vec{t} \ ,
$$ 
defines a complex Hermitian structure on the 
tangent space of $S^2$ at $\vec{S}$. In what follows, we shall regard   
the tangent bundle of $S^2$ as a Hermitian (complex) line bundle, 
denoted by ${\bf T}_{S^2}$, with $SO(3)$ as the 
 unit sphere bundle. The map, $\vec{t} \mapsto 
\vec{t}_-$, defines a $\CZ$-linear isomorphism between 
${\bf T}_{S^2}$ and  
the complex line subbundle of 
${\bf T}_{S^2} \otimes_{\RZ} \CZ$ 
generated by $\vec{t}_-$.

Let $X$ be a simply-connected marked  
manifold. An $SU(2)$-connection $J$ on $X$ is expressed by  
 $$ 
J =   
 2 Im(q)e^1 - 2  Re(q)e^2 +  V e^3  =  
q \sigma^- - \overline{q}\sigma^+ 
+ V e^3   
$$
with $ q = q_{\mu}dx^\mu \in \Omega^1(X)_{\CZ} , \ 
 V= V_{\mu}dx^\mu \in \Omega^1(X) $. The associated $SO(3)$-connection 
$J'$ is given by
$$
J' =  (e^1, e^2, e^3)
 \left( \begin{array}{ccc}
          0 &V & 2Re(q)\\
          - V & 0&2 Im(q )\\
          -2Re(q)  &-2 Im(q) & 0
         \end{array}   \right)
\left( \begin{array}{c}
          e_{1} \\
          e_{2} \\
          e_{3}
         \end{array}   \right) \ . 
$$
It is known that the zero curvature of the connection $J$, $F(J)=0$, 
is the consistency condition of the linear 
system :
\bea(l)
d g = g (q \sigma^- - \overline{q}\sigma^+ 
+ V e^3 ) \ , \ \ g \in {\cal C}^\infty(X,SU(2)) \ , 
\elea(gJ)
which is equivalent to the system (\req(Adeq)), now  
in terms of $\vec{S}, \vec{t}$,  expressed by 
\bea(l)
\left\{ \begin{array}{lll}
d \vec{S} & = q \vec{t}_{-} + \overline{q} \vec{t}_{+} &  \\
d_A \vec{t}_+ & =-2 q \vec{S} , &{\rm for } \ \ A = - i V  \ .
\end{array} \right.
\elea(SU2Eq) 
By Proposition 1, the 1-forms $q, A$, satisfy the 
relation (\req(0F)), which is equivalent to the expression:
\bea(l)
\left\{ \begin{array}{ll}
F(A) = 2 q  \overline{q}  ,    & \\
d_A q = 0 &{\rm for } \ , \ \ A = - i V \ .
\end{array} \right.
\elea(SU2F)
The curvature $F(A)$ is related to 
$\vec{S}$ by the formula:
\bea(l)
\frac{1}{4\pi}\vec{S} \cdot (d\vec{S} \times d \vec{S}) = 
\frac{i}{\pi}  q \overline{q}  = \frac{i}{2\pi} F(A) \ .
\elea(ChernS)
In the case of a 2-dimensional Riemannian manifold $X$, one has 
the inequality:
$$
dS \cdot * dS \geq \pm \vec{S} \cdot (d\vec{S} \times d \vec{S}) \ \ , 
\ \ ( \ \Leftrightarrow \ \ \ 
 4 q ( * \bar{q}) \geq \pm 4i q \bar{q} \ ) \ .
$$
The following lemma is implied by Proposition 1.
\par \vspace{0.2in} \noindent
{\bf Lemma 2. } Through the relation (\req(SU2Eq)), 
there is an one-to-one correspondence between the following 
data for a simply-connected marked manifold $X$:

(I)$(\vec{S}, \vec{t}) \in {\cal C}^\infty ( X 
, SO(3))$. 

(II) $( A, q )$, a solution of (\req(SU2F)) with $ A \in i \Omega^1 (X) \ ,
q \in \Omega^1(X)_{\CZ} $.    
$\Box$ \par \vspace{0.2in} \noindent 
For $X = \RZ^2$, the local expressions of (\req(SU2Eq)), (\req(SU2F)),
are:
\bea(l)
\left\{ \begin{array}{ll}
\partial_\mu \vec{S} = q_\mu \vec{t}_- + \overline{q_\mu} \vec{t}_+ \ , 
&  \\
( \partial_\mu - i V_\mu ) \vec{t}_+ = - 2 q_\mu \vec{S} \ , &
\mu = 0 , 1 \ ,
\end{array} \right. \ \ 
\left\{ \begin{array}{l}
 \partial_0V_1 - \partial_1V_0  = 
 2 i ( q_0 \overline{q_1} - \overline{q_0} q_1 ) \ ,   \\
( \partial_0 -iV_0 ) q_1 = ( \partial_1-iV_1 ) q_0  \ .
\end{array} \right.
\elea(SU2FR2)
Let $H$ be the Cartan subgroup consisting of 
diagonal elements of $SU(2)$. As $H$ is the isotropic subgroup 
of the element $e^3$ for the adjoint action of $SU(2)$ on ${\bf g}$,   
$$
H  = \{ g \in SU(2) 
\ | \ {\rm Ad}(g) (e^3) = e^3 \} \ ,
$$
the homogeneous space    
$SU(2)/ H $ can be identified with the unit sphere $S^2$ 
in ${\bf g}$  with the base element $e= e^3 $. 
Note that the exponential 
of an element $i\xi \in i\RZ $ (= the Lie algebra of $H$) acts on ${\bf g}$ by
$$
(e^1, e^2, e^3) \mapsto (e^1, e^2, e^3) \left( \begin{array}{ccc}
          \cos 2 \xi &\sin 2\xi & 0\\
          -\sin 2\xi  & \cos 2 \xi&0\\
          0  & 0&1
         \end{array}   \right) \ .
$$
For $h \in {\cal C}^\infty(X, H)$, we have 
$$
h^{-1}dh =  ( d \alpha ) e^3 \ \ \ {\rm for}  \ \ \alpha \in  
\Omega^0(X) \ .
$$
Under the action of $h$, 
the transformations of $\vec{S}, \vec{t}, V, q$, in 
(\req(SU2Eq))  are given by  
\bea(ll)
\vec{S}  \mapsto  \vec{S} \ , &
\vec{t}_-  \mapsto  {\rm e}^{-i \alpha } \vec{t}_- \ , \\
q  \mapsto  {\rm e}^{i \alpha} q ,  & 
V  \mapsto  V +  ( d \alpha ) \ .
\elea(tran)
Consider $A \ (:= -iV)$ as an $U(1)$-connection on the trivial 
complex line bundle ${\bf L}$ over $X$ with the canonical Hermitian 
structure,  
$q$ as a section of 1-forms with values in ${\bf L}$. 
 There is a covariant differential 
$d_A$ induced by $A$ on ${\bf L}$-valued differential forms. The 
transformations of $q, V$, in 
(\req(tran)) are described by a different choice of trivialization 
of the line bundle ${\bf L}$. 
By Proposition 2, we have the following result: 
\par \vspace{0.2in} \noindent
{\bf Proposition 3.} Let $X$ be a simply-connected marked manifold with 
${\rm H}^2(X, \ZZ) = 0$, ${\bf L}$ be the trivial complex line 
bundle over $X$. There is an one-to-one correspondence of the 
following data:

(I) $\vec{S} \in {\cal C}^\infty (X, S^2)$.

(II) $(V, q )$, $V \in \Omega^1 (X), 
q \in \Omega^1(X)_{\CZ} $, a solution of  (\req(SU2F)) modulo the 
relation $(V, q ) \sim 
( \tilde{V}, \tilde{q} ) $: 
$$
( \tilde{V}, \tilde{q})  = ( V +  ( d \alpha ) , {\rm e}^{i \alpha} q  ) , \
\ {\rm for  \ } \alpha \in \Omega^0(X) \ .
$$

(III) $(A, s) $, where $A$ is an $U(1)$-conncetion on ${\bf L}$,   
$s \in \Omega^1(X , {\bf L})$, satisfying the following  relation:
\bea(l)
\left\{ \begin{array}{l}
F(A) =  2 s s^{\dag}  ,    \\
d_A (s) = 0 . 
\end{array} \right.
\elea(SU2Fi)
The above $\vec{S}$ , $(V, q )$, are related by  
(\req(SU2Eq)) for some $\vec{t} \in {\cal C}^\infty(X, S^2)$, and 
$( A, s)$, $(V, q)$, by  
the relation: $(A, s) = ( -iV, q)$,   
via a trivialization of ${\bf L}= X \times \CZ$ ( as $U(1)$-bundles ).  
$\Box$ \par \vspace{.2in} \noindent
Note that    
the vector-valued 1-form $q \vec{t}_-$ in the above proposition 
is independent of 
the choice of gauge transformation $\alpha$ in (II), hence it defines the section 
$s$ of (III). The Hermitian line bundle ${\bf L}$ is isomorphic to 
the pull-back of line bundle ${\bf T}_{S^2}$ via the map 
$\vec{S}$. Therefore one can formulate the conclusion   
of Proposition 3 in the following instrinsic form.  
\par \vspace{.2in} \noindent 
{\bf Theorem 1. } Let ${\bf L}$ be a Hermitian line 
bundle over a simply-connected marked manifold $X$. There is an one-to-one correspondence of the 
following data:

(I) $\vec{S} \in {\cal C}^\infty (X, S^2)$ with $\vec{S}^*({\bf T}_{S^2}) 
= {\bf L}$ .

(II) $(A, s)$, a solution of (\req(SU2Fi)), where $A$ is an $U(1)$-connection on ${\bf L}$, 
$s \in \Omega^1(X , {\bf L})$. 
\par \noindent
In fact, $( A, s)$ in (II) is expressed by 
$( -iV, q)$ 
via a  trivialization of ${\bf L}$ over a local open 
set $U$, which is related to the $\vec{S}$ in (I)  
by (\req(SU2Eq)) for some $\vec{t} \in {\cal C}^\infty(X, S^2)$.   
Furthermore, the above correspondence satisfies the following 
functorial property.  Namely, let $X_j$, $ ( j= 1,2 )$, be two 
simply-connected marked manifolds with the corresponding elements,    
$(A_j, s_j)$ in (II),     
$\vec{S}_j \in {\cal C}^\infty (X_j, S^2)$ in (I). Assume 
$\vec{S}_1 = \vec{S}_2 \varphi $ for some 
marked map $\varphi: X_1 \longrightarrow X_2$. Then $(A_1, s_1) = ( \varphi^*(A_2), 
\varphi^*(s_2))$.
\par \vspace{.2in} \noindent
{\it Proof.} Let $\{ U_\alpha \}$ be a simple open cover of $X$, 
$u_{\alpha}$ be a base element  of $ U_\alpha$. By 
Proposition 3, an element $(A, s)$ of (II) gives rise 
a collection of maps,  
$\vec{S}_{\alpha} : (U_{\alpha}, u_{\alpha} ) \longrightarrow 
( S^2 , e ) $, with $\vec{S}_{\alpha}^*(({\bf T}_{S^2}) = 
{\bf L}|_{U_{\alpha}}$, and vice versa. 
For each pair $(\alpha, \beta)$, there is an unique element $g_{\alpha, \beta}$ 
in $SO(3)$ such that 
$$
\vec{S}_{\alpha}(x)= g_{\alpha, \beta}(\vec{S}_{\beta}(x)) \ , \ \ 
{\rm for} \  x \in U_{\alpha} \cap U_{\beta} \ .
$$
Then $\{ g_{\alpha, \beta} \}$ forms a cocycle for the cover 
$\{ U_\alpha \}$ with values in $SO(3)$. By the simply-connectness of 
$X$, there exists an 1-cochain $\{ g_\alpha \}$ such that 
$g_{\alpha, \beta} = g_\alpha^{-1} g_\beta$.  
Then the function, 
$$
\vec{S}(x) : = g_\alpha (\vec{S}_{\alpha}(x)) \ ,  
$$ 
is the marked map from $X$ to $S^2$ in (I). The 
functorial property follows easily from Proposition 3.  
$\Box$ \par \vspace{.2in} \noindent
To illustrate the correspondence of Theorem 1, we are going to give 
the explicit construction of a solution of (II) for 
the complex 
projective plane ${\PZ}^1$, corresponding to the 
canonical identification of $S^2$  with the extended complex plane via the 
stereographic projection. 
\par \vspace{.2in} \noindent
{\bf Example.} The Fubini-Study metric on ${\PZ^1}$, which is identified    
with the one-point compactification of the complex plane, 
\bea(l)
{\PZ}^1 = \CZ \cup \{ \infty \} \ , \ 
[1, 0 ] \leftrightarrow \infty, \
 [ \zeta, 1 ] \leftrightarrow \zeta  \in \CZ \ .
\elea(P1) 
The stereographic projection defines the diffeomorphism,  
\bea(ll)
p : \PZ^1 \longrightarrow S^2 \ , \ \ \zeta \mapsto \vec{S}  \ , \ {\rm with } \ 
\ \ S_1 + i S_2 = \frac{2 \zeta}{1+|\zeta|^2} \ , \ &  \ 
S_3 = \frac{1-|\zeta|^2}{1+|\zeta|^2} \ ,
\elea(stereo)
which sends the origin, $\zeta =0$, to the base element of $S^2$.  
The Fubini-Study metric on ${\PZ}^1$ is a  
metric invariant under fractional tranformations of $SU(2)$ on 
$\PZ^1$,
\bea(l)
\zeta \mapsto \zeta' = 
\frac{\alpha \zeta + \beta}{\gamma \zeta + \delta} \ , \ \ \ \  
\left( \begin{array}{cc}
          \alpha & \beta \\
          \gamma &\delta
         \end{array}   \right) \in SU(2) \ . 
\elea(mtraf)
In the coordinate $\zeta$,  the metric is expressed by 
$$
 \frac{d \zeta  \otimes d \overline{\zeta}}
{(1+|\zeta|^2)^2 } \ , 
$$ 
with the 2-form,  
$$
(1+|\zeta|^2)^{-2} d \zeta d \overline{\zeta} = - 
\overline{\partial} {\partial} \log (1+|\zeta|^2) \ .
$$
Furthermore, the expressions in the above hold also for any coordinate 
system $\zeta'$ related to $\zeta$ by (\req(mtraf)). With the canonical 
metric on $S^2$ and the Riemannian metric $
4(1+|\zeta|^2)^{-2} Re ( d \zeta  \otimes d \overline{\zeta} ) 
$ on ${\PZ^1}$, the map (\req(stereo)) is an $SU(2)$-equivariant 
isometry. One can set $\vec{t}_{\pm}$ on the $\zeta$-plane by 
$$
\vec{t}_-=  (1+|\zeta|^2) \partial_{\zeta} \vec{S} \ , \ \ \ 
\ \vec{t}_+ = (1+|\zeta|^2) \partial_{\bar{\zeta}} \vec{S} \ , 
$$
hence 
$$
\vec{S} = \frac{1}{1+|\zeta|^2}\left( \begin{array}{c}
           \zeta + \bar{\zeta} \\
          -i ( \zeta - \bar{\zeta})\\
          1- |\zeta|^2
         \end{array}   \right) \ , \ \ 
\vec{t}_+  = 
\frac{1}{1+|\zeta|^2}\left( \begin{array}{c}
           1- {\zeta}^2 \\
          i(1 +{\zeta}^2)\\
          -2 \zeta
         \end{array}   \right) \ , \ \
\vec{t}_-  = 
\frac{1}{1+|\zeta|^2}\left( \begin{array}{c}
           1- \bar{\zeta}^2 \\
          -i(1 + \bar{\zeta}^2)\\
          -2 \bar{\zeta}
         \end{array}   \right) \ . 
$$
By computation, one obtains the expressions of $q, A$, in (\req(SU2Eq)), 
denoted by $\breve{q}, \breve{A}$, as follows:
$$
\breve{q} = \frac{d \zeta}{1 + |\zeta|^2} \ , \ \ 
\breve{A} = - \partial \log (1+ |\zeta|^2) + \overline{\partial}
\log (1+ |\zeta|^2) \ . 
$$ 
By the $SU(2)$-invariant properties of metrics on ${\PZ^1}$ and 
$S^2$, the expressions of $\breve{q}$, $\breve{A}$, in the above 
formula hold 
also for any coordinate system $\zeta'$ in (\req(mtraf)).  
The connection $\breve{A}$ is characterized as an $SU(2)$-invariant 
$U(1)$-connection on the tangent bundle of $\PZ^1$. 
Note that the $U(1)$-bundle $p^* {\bf T}_{S^2}$ is the anti-canonical 
bundle ${\bf K}^{-1}$ over ${\PZ^1}$ endowed with the Fubini-Study metric. 
The local frame for  Eq.(\req(SU2Eq)) is 
$(1+ |\zeta'|^2)^{-1} 
\frac{\partial}{\partial \zeta'}$. Hence the element $(A, s)$  
in Theorem 1 is given by the $SU(2)$-invariant 
connection $\breve{A}$ and 
the "tautological" section $\breve{\Phi}$, 
\bea(cll)
d_{\breve{A}} ( (1+ |\zeta'|^2)^{-1} 
\frac{\partial}{\partial \zeta'}) & = &
(- \partial \log (1+ |\zeta'|^2) + \overline{\partial}
\log (1+ |\zeta'|^2) ) (1+ |\zeta'|^2)^{-1}  
\frac{\partial}{\partial \zeta'} \ ,
\\
\breve{\Phi} & = & d \zeta' \frac{\partial}{\partial \zeta' } \ \in \ \Omega^{1,0}( \PZ^1, 
{\bf K}^{-1} ) \ .
\elea(Usol)
By the universal property of Theorem 1, the above $(\breve{A}, 
\breve{\Phi}) $ can be   
served as the universal object of solutions of (\req(SU2Fi)). Hence 
we obtain the following   
statement:  
\par \vspace{.2in} \noindent
{\bf Corollary.} The expression of the 
corresponding element $(A, s)$ in Theorem 1 (II) 
for an element $\vec{S} \in {\cal C}^\infty (X, S^2)$
is given by 
$$
(A,s) = ( \zeta ^*(\breve{A}) , \zeta^*(\breve{\Phi}) ) \ , 
\ \ {\rm where } \ 
\zeta : = p^*(\vec{S}) : X \longrightarrow {\PZ}^1 \ .
$$
$\Box$ \par \vspace{.2in} \noindent

\section{Self-Dual Equation and Liouville Equation } 
In this section, $X$ will always denote a Riemann surface, 
i.e. an one-dimensional complex manifold, with  
a local coordinate system, $ z = x + i y $. With   
the holomorphic and anti-holomorphic differentials,
$$
d z = d x + i d y \in \Omega^{1,0}(X) , \ \ \ 
d \overline{z} = d x - i d y \in \Omega^{0, 1}(X) \ ,
$$
one decomposes differential forms of $X$ into 
$({\rm p, q})$-forms,  
$$
\Omega^*(X)_{\CZ} = \bigoplus_{\rm p, q} \Omega^{\rm p, q}(X) \ , 
$$
hence the differential $d = d' + d''$, 
$$
d' : \Omega^{\rm p, q}(X) \longrightarrow \Omega^{\rm p+1, q}(X) \ , \ \
d'' : \Omega^{\rm p, q}(X) \longrightarrow \Omega^{\rm p, q+1}(X) \ .
$$ 
For $q \in \Omega^1(X)_{\CZ}$, we have 
$$
q = q' + q''\ , \ \ \  q' \in \Omega^{1,0}(X) \ , \ \ 
q'' \in \Omega^{0,1}(X) \ ,
$$
with the local expressions,  
$$
q = q_{x} dx + q_{y} dy  \ , \ 
 \ q' = q_{z} d z  \ , 
\ \ \ q''= q_{\overline{z}} d \overline{z}  \ \ \ {\rm where} \ \ 
q_{z}  := 
\frac{q_{x} - i q_{y} }{2} \ , \ 
q_{\overline{z}} : =  \frac{q_{x} + i q_{y}}{2} \ .
$$
For an $U(1)$-connection $A \in i\Omega^1(X)$, 
one has the decomposition of $d_A$: 
$$
d_A = d'_A + d''_A , \ \ d'_A := d' + A' \ , \ \ d_A'':= d'' + A'' \ . 
$$
Note that $A'' = - \overline{A'}$. In a local 
coordinate $z$, we shall write 
$$
d_A' ( f ) = d z (\partial_A)_{z} f \ \ , \ 
d_A'' ( f ) = d \bar{z} (\partial_A)_{\bar{z}} f \ \ \  \ 
{\rm for }  \ \ f \in \Omega^0(U)_{\CZ} \ . 
$$
Furthermore one has the decomposition of the ${\bf L}$-valued forms,
 where ${\bf L}$ is a Hermitian line bundle over $X$ with an 
$U(1)$-conncetion $A$, 
$$
\begin{array}{ll}
\Omega^*(X, {\bf L}) = \bigoplus_{\rm p, q} \Omega^{\rm p, q}(X, {\bf L}) \ ,& 
d_A = d'_A + d''_A \ ,\\
d'_A : \Omega^{\rm p, q}(X, {\bf L}) \longrightarrow 
\Omega^{\rm p+1, q}(X, {\bf L}) \ , & 
d''_A : \Omega^{\rm p, q}(X, {\bf L}) \longrightarrow 
\Omega^{\rm p, q+1}(X, {\bf L}) \ .
\end{array}
$$
For $s \in \Omega^1(X, {\bf L})$, we denote 
$$
s = s' + s'' \ , \ \ \ 
s' \in \Omega^{1,0}(X, {\bf L}) \ , \ \ 
s'' \in \Omega^{0,1}(X, {\bf L}) \ .
$$
Hence  
$$
F(A) = d'' A' + d' A'' \ , \ \ \ 
d_A s = d_A'' s' + d_A' s'' \ , \ \ 
s s^{\dag} = s' s'^{\dag} - s''^{\dag}s'' \ .
$$
Locally we have the expressions,  
$$
A = A_z  d z + A_{\bar{z}} d \bar{z} \ , \ \ \ 
s = q_z  d z + q_{\bar{z}} d \bar{z} \ , \ \
$$
then (\req(SU2Fi)) becomes
\bea(l)
\left\{ \begin{array}{l}
\partial_z A_{\bar{z}} - \partial_{\bar{z}} A_z =  2 ( 
|q_z|^2 - |q_{\bar{z}}|^2)  ,    \\
(\partial_A )_z q_{\bar{z}} = (\partial_A )_{\bar{z}} q_z \ , 
\end{array} \right.
\elea(SU2Fhol)
which is the compactibility condition of the linear system (\req(gJ)), 
now taking the form:
$$
\left\{ \begin{array}{l}
d'g = g ( q' \sigma^- - \overline{q''} \sigma^+ 
+ V' e^3 ) \ ,    \\
d'' g = g (q'' \sigma^- - \overline{q'}\sigma^+ 
+ \overline{V'} e^3 ) \ ,
\end{array} \right. 
$$
with $ g \in {\cal C}^\infty(X, SU(2))$. An equivalent 
expression is the system: 
\bea(l)
\partial_z g = g \left( \begin{array}{ll}
\frac{i}{2} V_z &  - \overline{q_{\bar{z}}} \\
q_z & \frac{-i}{2} V_z
\end{array} \right) \ , \ 
\partial_{\bar{z}} g =  g \left( \begin{array}{ll}
\frac{i}{2} \overline{V_z}  & - \overline{q_z}  \\
q_{\bar{z}}  & \frac{-i}{2}\overline{V_z}
\end{array} \right) \ . 
\elea(Ccon)
Note that the r.h.s. in the above is a $SL_2(\CZ)$-connection.

Consider the following constraint, called 
the self-dual equation, of   
functions $\vec{S}$ from a marked Riemann surface $X$ to $S^2$:
\bea(lll)
d' \vec{S} = i  \vec{S} \times d' \vec{S} \ ,  &
( \Leftrightarrow \ &
 d' \zeta = 0 \ ) ,  
\elea(sSz)
or
\bea(lll)
d'' \vec{S} = i  \vec{S} \times d'' \vec{S} \ &
( \Leftrightarrow \ &
d'' \zeta = 0  \ ) ,  
\elea(sSzz)
where $\zeta$ is  
the composition of $\vec{S}$ with the stereographic projection  
(\req(stereo)), $\zeta : X \longrightarrow \PZ^1$.
By the formula of vector product in $\RZ^3$, 
$$
v_1 \times ( v_2 \times v_3 ) = ( v_1 \cdot v_3) v_2 - 
( v_1 \cdot v_2 ) v_3 \ , 
$$
Eqs. (\req(sSz)), (\req(sSzz)), have the following local expressions: 
$$
\begin{array}{lll}
(\req(sSz)) &\Leftrightarrow & 
 \partial_{x} \vec{S} =  \vec{S} \times \partial_{y}
 \vec{S} \ ,  \\
(\req(sSzz)) &\Leftrightarrow & 
 \partial_{x} \vec{S} =  - \vec{S} \times \partial_{y}
 \vec{S} \ .
\end{array}
$$
For a solution $\vec{S}$ of (\req(sSz)), let    
${\bf L}$ be the Hermitian line bundle $\vec{S}^*{\bf T}_{S^2}$, 
 $(A, s)$ be the solution of (\req(SU2Fi)) corresponding to 
$\vec{S}$ in Theorem 1. With a local 
trivilization of ${\bf L}$ over an open set $U$, $s$ is represented by 
an 1-form $q \in \Omega^1(U)_{\CZ}$. 
By (\req(SU2Eq)),  Eq.(\req(sSz)) is equivalent to 
the relation, $q' = 0 $, i.e. $
q_{x} =  i q_{y} $, and the same for      
Eq. (\req(sSzz)) and $q'' = 0$.  Hence  
$$
\begin{array}{lll}
(\req(sSz)) & \Leftrightarrow & s' = 0 \ , \ {\rm i.e} \ \ 
s = s''  \ , \ \\ 
(\req(sSzz)) & \Leftrightarrow &  s''= 0 \ , \ {\rm i.e} \ \ 
s = s' \ .
\end{array}
$$
Replace ${\bf L}, A,$ by the complex 
conjuate ${\bf L}^{\dag}, A^{\dag}$, in the case of (\req(sSz)), 
and set 
\[ 
 \Phi  = \left\{ \begin{array}{ll}
  s^{\dag}  & {\rm for } \ \ (\req(sSz)) \ , \\
 s  & {\rm for } \ \ (\req(sSzz)) \ .
\end{array} \right.
\] 
We have 
$$
d_A \Phi = d''_A \Phi \ , 
$$  
and Eq.(\req(SU2Fi)) becomes 
\bea(l)
\left\{ \begin{array}{l}
F(A) = 2 \Phi \Phi^*  \ , \\  
d''_A \Phi = 0     \ ,
\end{array} \right.
\elea(selfdual)
where $A$ is an $U(1)$-connection of a Hermitian  
line bundle ${\bf L}$ over $X$, and 
$\Phi \in \Omega^{1,0}(X, {\bf L})$. By Theorem 1, 
 we obtain the following result:
\par \vspace{.2in} \noindent 
{\bf Theorem 2. } Let ${\bf L}$ be a Hermitian complex line 
bundle over a simply-connected marked Riemann surface $X$. There is an 
one-to-one correspondence between the following data:

(I) $\vec{S} \in {\cal C}^\infty (X, S^2)$, a 
solution of (\req(sSz)) with $\vec{S}^*({\bf T}_{S^2}^{\dag}) 
= {\bf L}$ .

(II) $\vec{S} \in {\cal C}^\infty (X, S^2)$, a 
solution of (\req(sSzz)) with $\vec{S}^*({\bf T}_{S^2}) 
= {\bf L}$ .

(III) $(A, \Phi)$, a solution of Eq.   
(\req(selfdual)), where $A$ is an $U(1)$-connection on ${\bf L}$, 
$\Phi \in \Omega^{1,0}(X , {\bf L})$. 
\par \noindent
The $\vec{S}$'s in (I), (II), are related by the transformation, 
$$
\vec{S} \leftrightarrow \left( \begin{array}{ccc}
          1&0&0 \\
          0&-1&0\\
          0&0&1
         \end{array}  \right)\vec{S} \ . 
$$
For $(A, \Phi)$ of (III), the corresponding $\vec{S}$ in (I) ( or (II)) 
is the element related to $(A^{\dag}, \Phi^{\dag})$ ( or 
$(A, \Phi)$ respectively ) in Proposition 3 (I). 
$\Box$ \par \vspace{.2in} \noindent
{\bf Remark.} (i) Replacing  $X$ of the above theorem 
by an arbitrary Riemann surface,  by the relation, 
$$
d_A \Phi = 0 \Longrightarrow d_A'' \Phi = 0  \ , \ \ 
{\rm for} \ \Phi \in \Omega^{1,0}(X, {\bf L}) \ ,
$$
one obtains the following relations among (I) (II) and (III),
$$
{\rm (I)} \Leftrightarrow {\rm (II)} \Longrightarrow {\rm (III)} \ .
$$   
(ii) 
The general solution of (II) in Theorem 2  
for a  simply-connected marked Riemann surface $X$ (with a base element $x_0$) is given by 
$$
(A , \Phi ) = (\varphi^*(\breve{A}), \varphi^*(\breve{\Phi})) \ , \ \ 
\varphi: (X , x_0) \longrightarrow (\PZ^1, 0) \ \ {\rm holomorphic} \ ,
$$
with ${\bf L} = \varphi^*({\bf K}^{-1})$. Here $(\breve{A}, 
\breve{\Phi})$ is the solution over ${\PZ}^1$ described by (\req(Usol)). 
 For $X = \PZ^1$, the  
map $\varphi$ is a rational function of 
$X$. By (\req(ChernS)), 
the Chern class of 
$\varphi^*({\bf K}^{-1})$ is given by
$$
c_1(\varphi^*({\bf K}^{-1})) = 
\frac{1}{4\pi}\int \vec{S} \cdot (d\vec{S} \times d \vec{S}) = 
2 \ ( {\rm degree \ of \ } \varphi) \ .
$$
$\Box$ \par \vspace{.2in} \noindent
{\bf Corollary. } For $X = \CZ$ with the coordinate $z = x + i y$, 
let ${\bf L}$ be the trivial line bundle over $X$ with the canonical 
Hermitian structure. Then   
there is an one-to-one correspondence between the 
following data:

(I) $\vec{S} : X \longrightarrow S^2  $, a solution of (\req(sSz)) 
( or (\req(sSzz)) respectively ) with $\vec{S}(0,0)= e$. 

(II) $(A, \Phi)$, a solution of (\req(selfdual)) where $A$ is an  
$U(1)$-connection of ${\bf L}$, $\Phi \in \Omega^{1,0}(X, {\bf L})$.

(III) $\phi : X \longrightarrow \RZ \cup \{ - \infty \} $,  a solution of 
Liouville equation:
\bea(ll)
({\rm LE}) \ : &  
- (\partial_x^2 + \partial_y^2)  \phi  = 
8 {\rm e}^{ \phi} \  .
\elea(LE) 
The relation of (I), (II), is given by Theorem 2. The relation of  
(II), (III), is given  by 
\bea(ll)
A = A_z d z + A_{\bar z} d {\bar z} \ , \ & 
A_{\bar{z}}  =  \frac{-1}{2} \partial_{\bar{z}}\phi  \ , \\
\Phi = {\rm e}^{\frac{ \phi}{2}} dz \ , 
\elea(LEphi)
and (I), (III), by 
$2 {\rm e}^{\phi} = |\partial_z \vec{S}|^2$   
( or $ | \partial_{\bar z} \vec{S}|^2 $ respectively ). 
\par \vspace{.2in} \noindent
{\it Proof.} As $X$ is a contractable space, we may assume the line 
bundle ${\bf L}$ in Theorem 2 is the trivial one. Hence it follows 
the relation between (I), (II). For an element $(A, \Phi)$ 
of (II), by a suitable frame of ${\bf L}$ outsider the   
zeros of $\Phi$, one can write 
$\Phi = {\rm e}^{\frac{ \phi}{2}} dz$ for a function $ 
\phi$ with values in $\RZ \cup \{ - \infty \}$. Then 
Eq. (\req(selfdual)) is equivalent to the relation,  
$A_{\bar{z}}  =  \frac{-1}{2} \partial_{\bar{z}} \phi$. Hence 
$\phi$ satisfies Eq. (\req(LE)) of (III). The relation between 
(I), (III), follows easily from  (\req(SU2FR2)).
$\Box$ \par \vspace{.2in} \noindent
{\bf Remark.} By Corollary of Theorem 1 and 
the above relation between (II), (III), one obtains 
the well-known expression of a general solution of Liouville equation:
$$
\phi ( z ) =  \log \frac{| \partial_z \zeta (z)|^2 }{(1 + | \zeta (z) |^2)^2} \ \ 
, \ \ \ \zeta: \CZ \longrightarrow 
\PZ^1 \ \ {\rm holomorphic}  \ . 
$$

\section{Conformal Sinh-Gordon Equation}
In this section, $X$ is a simply-connected marked Riemann surface.  
We are going to discuss the following equation of 
$\vec{S} \in {\cal C}^\infty(X, S^2)$,  
\bea(l)
 \vec{S} \times d''d' \vec{S} = 0 \ , \ \ 
( \ \Longleftrightarrow \ \ 
 d''d' \vec{S} + (d'' \vec{S} 
, d' \vec{S} ) \vec{S} = 0  \ \Longleftrightarrow \ \ 
d'd''\zeta =   
\frac{2\bar{\zeta}}{1+ |\zeta|^2}d'\zeta d''\zeta \ ) \ ,
\elea(SGS)
where $\zeta = p^*(\vec{S})$.
The local expression of (\req(SGS)) is given by
\bea(l)
 \vec{S} \times ( \partial_x^2 + \partial_y^2 ) \vec{S} = 0 \ , \ \ 
( \ \Longleftrightarrow \ \ 
 \partial_{\bar z} \partial_z \vec{S} + (\partial_{\bar z} \vec{S} 
+ \partial_z \vec{S} ) \vec{S} = 0  \ \Longleftrightarrow \ \ 
\partial_{z}\partial_{\bar z} \zeta = 2{\bar \zeta} 
\frac{\partial_{z} \zeta \partial_{\bar z} \zeta}{1+ |\zeta|^2} \ ) \ .
\elea(SGSloc)
Let  $(A, s)$ be 
the solution of (\req(SU2Fi))  corresponding to $\vec{S}$ in 
Theorem 1. Eq.(\req(SGS)) gives rise a constraint of $(A, s)$, which is 
the expressions, 
$$
d_A''s' + d_A's'' = d_A''s' - d_A's'' =  
d_A''s' =  d_A's'' = 0 \ \ .
$$
Note that the zeros of any two of the above four sections imply 
the others. Locally they are described by the relations: 
$$
(\partial_A)_x q_y - (\partial_A)_y q_x = 
(\partial_A)_xq_x + (\partial_A)_yq_y = 0  \ . 
$$
Obviously these conditions are preserved under  
holomorphic transformations of the Riemann surface $X$. 
\par \vspace{.2in} \noindent 
{\bf Theorem 3. } Let ${\bf L}$ be a Hermitian line  
bundle over a simply-connected marked Riemann surface $X$. Under the 
correspondence of Theorem 1, the following data are equivalent:

(I) $\vec{S} \in {\cal C}^\infty (X, S^2)$, a 
solution of (\req(SGS)) with $\vec{S}^*({\bf T}_{S^2}) 
= {\bf L}$ .

(II) $(A, s)$, where $A$ is a $U(1)$-connection on ${\bf L}$, 
$s \in \Omega^1(X , {\bf L})$, satisfying the equation  
\bea(l)
\left\{ \begin{array}{l}
F(A) =  2  s s^{\dag} \ , \\
d_A''s' = d_A's'' = 0  \ .
\end{array} \right.
\elea(SGF)
Furthermore, the $\vec{S}$'s of (I) corresponding to those elements 
of (II) with  $s'=0$ ( or $s''=0$ ) 
are solutions of (\req(sSz)) ( or (\req(sSzz)) respectively ) 
 described by Theorem 2 before.
$\Box$ \par \vspace{.2in} \noindent
Now we are going to establish the relation between Eq.(\req(SGF)) and 
the (conformal) sinh-Gordon equation:
\bea(lll)
({\rm shG})_u : & - (\partial_x^2 + \partial_y^2)   \phi  = 
8 ( u {\rm e}^{ \phi} -   {\rm e}^{- \phi} )  \ , &
\phi : \CZ \longrightarrow \RZ \cup \{ - \infty \} \ ,
\elea(sinhGg) 
where $u = |U|^2$ for a (fixed) holomorphic function $U$ of the 
complex plane $\CZ$. For $u = 0$, Eq. (\req(sinhGg)) is  
the Liouville equation, which posseses the conformal invariant property. 
Our main discussion here will be on the case of a non-zero $u$. 
Note that by Corollary of Theorem 1, the function,  
$$
\phi ( z ) =  \log \frac{| \partial_z \zeta (z)|^2 }{(1 + | \zeta (z) |^2)^2} \ \ 
\ \ \ {\rm with} \ \  
\partial_{z}\partial_{\bar z} \zeta = 2{\bar \zeta} 
\frac{\partial_{z} \zeta \partial_{\bar z} \zeta}{1+ |\zeta|^2} \ \ ,
$$  
is a solution of (\req(sinhGg)) for 
$$
u = |U|^2 \ , \ \ 
U = \frac{\partial_z{\zeta} \partial_z \bar{\zeta}}{(1+ |\zeta|^2)^2} \ . 
$$
\par \vspace{.2in} \noindent 
{\bf Proposition 4. } Let $X = \CZ$ with the coordinate $z = x + i y$, 
and ${\bf L}$ be the trivial line bundle over $\CZ$ endowed with the 
natural Hermitian structure. Let ${\bf K}$ be the canonical bundle over 
$X$ with the vector space of holomorphic 
sections, $\Gamma ( X, {\bf K})$. Then   
there is an one-to-one correspondence between the 
following data:

(I) $\vec{S} : \CZ \longrightarrow S^2  $, a solution of 
(\req(SGSloc)), but not a solution of (\req(sSzz)), 
 with $\vec{S}(0)= e$. 

(II) $(A, s)$, a solution of (\req(SGF)), where $A$ is a 
$U(1)$-connection of ${\bf L}$, $s \in \Omega^1(\CZ, {\bf L})$ 
with $s'' \neq 0$.

(III) $(h , \sigma)$, where $\sigma \in \Gamma ( X, {\bf K})$,   
$h = <dz, dz>$ for a Hermetian metric $< , >$ of ${\bf K}$,  satisfying  
the relation: 
\bea(l)
d'' d' \log h = h \sigma \sigma^{\dag} -  h^{-1} 
dz d \bar{z} \ .
\elea(SGFhol)

(IV) $(\phi, U)$, where $U$ is an entire function,  
$\phi$ is a solution of (\req(sinhGg)) with 
$u = |U|^2 $. \par \noindent  
The correspondence between (I), (II), is given in Theorem 1. 
The relations between  
(II), (III), (IV), are described  by 
$$
\begin{array}{cl}
A =\frac{1}{2} ( h^{-1} d'h - h^{-1} d''h) \  , & 
s = h^{\frac{1}{2}} \sigma +  h^{- \frac{1}{2}} d\bar{z} \ , \\
A = \frac{\partial_{z}\phi}{2}  dz - 
\frac{\partial_{\bar{z}}\phi}{2}  d\bar{z} \ , & 
s = U {\rm e}^{\frac{\phi}{2}} dz + 
{\rm e}^{-\frac{\phi}{2}} d \bar{z} \ , \\
h = {\rm e}^{\phi} \ , & \sigma = U dz \ .
\end{array}
$$ 
Furthermore, the above correspondences are equivariant under the 
following actions of analytic automorphisms $f$ of $\CZ$ with 
$f(0) = 0$: 
$$
f^*\vec{S} \longleftrightarrow  (f^*A, f^*s) 
\longleftrightarrow ( |\partial_zf|^2 f^*h, f^*\sigma ) 
\longleftrightarrow  (\tilde{\phi}, \tilde{U}) \ , 
$$ 
where $\tilde{\phi}, \tilde{U}$, are defined by 
\bea(l)
\tilde{\phi}(z) = \phi(f(z)) + 2 \log |\partial_z f(z)| , \ \ \ \ 
\tilde{U}(z ) =  U(f(z)) (\partial_z f(z))^2 \ .
\elea(sinhchg)   
\par \vspace{.2in} \noindent
{\it Proof.} One may assume the line 
bundle ${\bf L}$ in Theorem 3 to be the trivial one, hence  
the relation between (I) and (II) follows immediately . 
We are going to show the equivalence of (II) and (IV). In terms of 
a suitable frame of ${\bf L}$, 
the element $(A, s)$ of (II) is expressed by
$$
s = q_z dz + q_{\bar{z}} d \bar{z} \ , \ \ 
A = A_z dz + A_{\bar{z}} d \bar{z} \ .
$$
Then (\req(SGF)) becomes
$$
\left\{ \begin{array}{l}
\partial_z A_{\bar{z}} - 
\partial_{\bar{z}} A_z =  2  ( | q_z|^2 - | q_{\bar{z}} |^2 ) \ ,   \\
(\partial_{\bar{z}} + A_{\bar{z}}) q_z = 
(\partial_z + A_z  ) q_{\bar{z}} = 0  \ .
\end{array} \right.
$$
Define 
$$
U  = q_z \overline{q_{\bar{z}}} \ .
$$
By the property $A_{\bar{z}} = - \overline{A_z}$, we have 
$$
\partial_{\bar{z}}U = (\partial_{\bar{z}} + A_{\bar{z}}) q_z = 
(\partial_z + A_z  ) q_{\bar{z}} = 0 \ .
$$
Since the vanishing of any two in the above implies the third one, one 
may replace $
( A , q_z , q_{\bar{z}})$ by $( A ,  U, q_{\bar{z}} )$. Then  
the equation becomes 
$$
\left\{ \begin{array}{l}
\partial_z A_{\bar{z}} - 
\partial_{\bar{z}} A_z =  2  ( | q_z|^2 - | q_{\bar{z}} |^2 )  \ , \\ 
\partial_{\bar{z}} U = (\partial_z + A_z  ) q_{\bar{z}} = 0  \ .
\end{array} \right.
$$
By a suitable unitary frame of ${\bf L}$, we may assume 
$q_{\bar{z}} = {\rm e}^{\frac{- \phi}{2}}$, where $ \phi: 
\CZ \longrightarrow \RZ $. 
Then we have 
$$
A_z =  \frac{1}{2} \partial_z \phi \ , 
$$
where $\phi$ is a solution of Eq. (\req(sinhGg)). Hence we obtain the 
correspondence between (II), (IV). The rest relations follow immediately. 
$\Box$ \par \vspace{.2in} \noindent
{\bf Remark.} With the same argument as in the above proposition, the
equivalent form of  
Eq. (\req(SGF)) with $s' \neq 0$ gives rise the equation 
of $(\phi, U)$ in \cite{BB}: 
$$
\left\{ \begin{array}{l}
- (\partial_x^2 + \partial_y^2)   \phi  = 
8 ( {\rm e}^{ \phi} - |U|^2 {\rm e}^{- \phi} )  \  , \\
\partial_{\bar{z}} U  = 0  \ .
\end{array} \right.
$$
$\Box$ \par \vspace{.2in} \noindent
By Proposition 4, one can derive the zero-curvature 
representation of the sinh-Gordon equation 
(\req(sinhGg)) as follows. 
Let $U$ be an entire function with $|U|^2= u$. Then the data 
in Proposition 4 (IV) for a fixed $\phi$ are given by 
the collection of $(\phi, \lambda U)$ 
with $\lambda \in \CZ, |\lambda|=1$, each of which corresponds 
to an element $(A, s)$ in (II), then by (\req(Ccon)), 
gives rise a $SU(2)$-connection with the  
linear system,
\bea(ll)
\partial_z g = g \left( \begin{array}{ll}
\frac{-1}{4}\partial_z\phi &  - {\rm e}^{\frac{-\phi}{2}} \\
\lambda U {\rm e}^{\frac{\phi}{2}} & \frac{1}{4}\partial_z\phi
\end{array} \right) \ , &
\partial_{\bar{z}} g =  g \left( \begin{array}{ll}
\frac{1}{4} \partial_{\bar{z}}\phi  & - \frac{1}{\lambda}\bar{U}{\rm e}^{\frac{\phi}{2}}  \\
{\rm e}^{\frac{-\phi}{2}}  & \frac{-1}{4} \partial_{\bar{z}}\phi
\end{array} \right) \ , 
\elea(zeroFshG)
for $|\lambda | = 1$, hence  $\lambda \in \CZ^*$ by extension. The 
variable 
$\lambda$ is called the spectral parameter. There is another 
intepretation of the spectral parameter by 
using the $*$-involuion of $\Omega^1(X)_{\CZ}$:
$$
* : \Omega^1(X)_{\CZ} \longrightarrow 
\Omega^1(X)_{\CZ}  \ , \ \ \  * dx = dy \ , \ \ * dy = - dx 
\ . 
$$
In fact, for a trivialization of ${\bf L}$ 
in Proposition 4, one has the identification, 
$\Omega^1(X, {\bf L}) = \Omega^1(X)_{\CZ}$. Then Eq.  
(\req(SGF)) becomes  
$$
\left\{ \begin{array}{l}
F(A) =  2  q \overline{q} \ ,  \\
d_A q = d_A (*q) = 0   \ ,
\end{array} \right.
$$
which is equivalent to the system for  
$\theta \in \RZ/2\pi \ZZ$, 
$$
\left\{ \begin{array}{l}
F(A) =  2  q_{\theta} \overline{q_{\theta}} \ , \\
d_A q_{\theta} = d_A (*q_{\theta}) = 0   \ ,
\end{array} \right.
$$
where $ q_{\theta} : = q \cos \theta  + (*q) \sin \theta $. The 
corresponding element of   
Proposition 4 (III) is given by 
$$
\begin{array}{llll}
(h, \sigma) =& ({\rm e}^{\phi}, U dz ) , 
& \longleftrightarrow & (A, q ) \ , \\
(h, \sigma_\theta) = & ({\rm e}^{\phi}, {\rm e}^{-2i \theta} U dz ) , 
& \longleftrightarrow & (A, q_\theta ) \ .
\end{array}
$$
The family $\{ q_\theta \}$ provides a role of the hidden symmetry of Eq.(\req(zeroFshG)). 
On the other hand, one can also associate the spectral 
parameter $\lambda$ with the symmetries of Eq.(\req(sinhchg)). 
For $u = 0$ , it is the Liouville equation, which possesses the  
holomorphic symmetries of $\CZ$. For $u \neq 0$ in 
$({\rm shG})_{u}$, the relation (\req(sinhchg)) determines the following  
holomorphic functions $f$ which preserve the function $u$ of the form 
$ u_k$:
$$
\left\{ \begin{array}{ll}
u(z) = u_k( z ) : = |z|^{2k} , & k \in \ZZ_{\geq 0} , \\
f(z) = f_\lambda (z) : = \lambda z \ , & 
\lambda \in \CZ^* \ , \ 
|\lambda| = 1   \ .
\end{array} \right.
$$
Hence one obtains the following statement as a corollary of 
Proposition 4:
\par \vspace{.2in} \noindent 
{\bf Proposition 5.} Let ${\bf L}, {\bf K}$ be the same as in 
Proposition 4. For a non-negative integer $k$, 
there is an one-to-one correspondence between the 
following data:

(I) $(h , \sigma)$, a solution of (\req(SGFhol)) with 
$\sigma \sigma^{\dag} = |z|^{2k} dz d\bar{z}$, modulo the relation 
$(h , \sigma) \sim (f_{\lambda}^*h , f_{\lambda}^*\sigma)$ with 
$|\lambda| = 1$.

(II) $\phi$, a solution of (\req(sinhGg)) for 
$u = u_k $. \par \noindent 
For a function $\phi$ in (II), the corresponding element in (I) 
is represented by    
$$
h = {\rm e}^{\phi} \ , \ \sigma = z^k dz \ .
$$ 
$\Box$ \par \vspace{.2in} \noindent
For Eq. (\req(sinhGg)) with $u = u_k$, a solution 
$\phi$ generates an one-parameter family of solutions as follows:  
$$
\phi ( z, \bar{z} ) \mapsto \phi_{\theta} ( z, \bar{z} ) 
: = \phi ( {\rm e}^{i\theta} z, {\rm e}^{-i\theta} \bar{z} ) \ , \ \ 
\theta \in \RZ/2\pi \ZZ \ .
$$ 
The linear system corresponding to $\phi_{\theta}$ is given by 
$$
\partial_z g = g \left( \begin{array}{ll}
\frac{-1}{4}\partial_z\phi_{\theta} &  - {\rm e}^{\frac{-\phi_{\theta}}{2}} \\
z^k {\rm e}^{\frac{\phi_{\theta}}{2}} & 
\frac{1}{4}\partial_z\phi_{\theta}
\end{array} \right) \ , \ \ \ 
\partial_{\bar{z}} g =  g \left( \begin{array}{ll}
\frac{1}{4} \partial_{\bar{z}}\phi_{\theta}  & 
- \bar{z}^k{\rm e}^{\frac{\phi_{\theta}}{2}}  \\
{\rm e}^{\frac{-\phi_{\theta}}{2}}  & \frac{-1}{4} 
\partial_{\bar{z}}\phi_{\theta}
\end{array} \right) \ \ .
$$
By Proposition 4 and the relation (\req(sinhchg)), using 
the transformation $f_{{\rm e}^{-i\theta}}$, one can change 
the above system into 
(\req(zeroFshG)) with $U= z^k$, $\lambda = {\rm e}^{-(k+2)i \theta }$.

Replacing the coordinate 
$(x, iy)$ by $(t, x) \in \RZ^2$ in previous arguments of this section, 
one is able to derive the sine-Gordon equation. In fact, 
we define the light-cone coordinate,
$$
t^+ = t + x \ , \ \ t^- = t - x \ ,
$$
with the differential operators,
$$
\partial_+ = \frac{1}{2} (\partial_t + \partial_x ) \ , \ \
\partial_- = \frac{1}{2} (\partial_t - \partial_x ) \ .
$$
The equation is defined by
\bea(l)
 \vec{S} \times ( \partial_t^2 - \partial_x^2 ) \vec{S} = 0 \ , \ 
( \ \Longleftrightarrow \ \ 
 \partial_+\partial_- \zeta = 2{\bar \zeta} 
\frac{\partial_+ \zeta \partial_- \zeta}{1+ |\zeta|^2}   \ ) \ \ , \ 
\zeta = p^*(\vec{S}) \  .
\elea(lSGSloc)
The constraint of the corresponding $(A, s)$ 
of Proposition 3 is given by
$$
d_A^-s^+ + d_A^+s^- = d_A^-s^+ - d_A^+s^- =  
d_A^-s^+ =  d_A^+s^- = 0 \ .
$$
Locally, $s$ is an 1-form $q$ which satisfies the 
relation,  
$$
(\partial_A)_t q_t  = (\partial_A)_x q_x \ .
$$
Then    
(\req(lSGSloc)) is related to the sine-Gordon equation:
\bea(ll)
({\rm sG})_W :  & \left\{ \begin{array}{l}
\partial_+ \partial_- \Phi = - 4 {\rm e}^W \sin \Phi \ , \\
\partial_+ \partial_- W = 0 \ . 
\end{array} \right.
\elea(sinG) 
Indeed, by the same argument as in 
Proposition 4 one can derive the following fact.
\par \vspace{.2in} \noindent 
{\bf Proposition 6. } Let ${\bf L}$ be the trivial line bundle over $\RZ^2$ 
with the natural  Hermitian structure. Then   
there is an one-to-one correspondence between the 
following data:

(I) $\vec{S} \in {\cal C}^\infty (\RZ^2, S^2)$, a 
solution of (\req(lSGSloc)) with $\vec{S}^*({\bf T}_{S^2}) 
= {\bf L}$ .

(II) $(A, s)$, where $A$ is a $U(1)$-connection on ${\bf L}$, 
$s \in \Omega^1(\RZ^2 , {\bf L})$, satisfying the equation  
\bea(l)
\left\{ \begin{array}{l}
F(A) =  2  s s^{\dag} \ , \\
(\partial_+ + A_+) s_- = (\partial_- + A_-) s_+ = 0 \ .
\end{array} \right.
\elea(lSGF)

(III) $(\Phi, r_+, r_-)$, where $r_\pm$ are non-negative functions 
on $\RZ^2$ with $\partial_- r_+ = \partial_+ r_-  = 0$,  
$\Phi $ a solution of 
(\req(sinG)) for $W = \log r_+ + \log r_- $

\par \noindent
The correspondence between (I), (II), is given in Theorem 1,  and  
 (II) , (III), by
$$
A = -i \partial_- \Phi dt^- \ , \ \ s = r_+ {\rm e}^{i \Phi} dt^+ 
+ r_- dt^- \ . 
$$
Furthermore, the correspondences are equivariant under 
action of the following automorphisms of $\RZ^2$:
$$
f : (t^+, t^- ) \mapsto ( \alpha(t^+), \beta(t^-) ) \ \ 
{\rm with } \ \ 
f(0,0) = (0,0 ) \ , \ \ \ \frac{d \alpha}{d t^+} \geq 0 \ , \ 
\frac{d \beta}{d t^-} \geq 0 \ ,
$$
via the relations:
\[
f^*\vec{S} \longleftrightarrow  (f^*A, f^*s) 
\longleftrightarrow  (\tilde{\Phi}, \tilde{r}_+ , \tilde{r}_- ) 
:= (f^*\Phi \ , \  \frac{d \alpha}{d t^+}\alpha^* r_+ \ , \
  \frac{d \beta}{d t^-}\beta^* r_- ) \ .
\]  
$\Box$ \par \vspace{.2in} \noindent
When $W$ is the constant, $\log m^2$, the 
linear system  
associated to (\req(sinG)) gives rise to the family with a spectral 
parameter $\lambda \in \CZ^*$ :
$$
\begin{array}{ll}
\partial_+ g = g \left( \begin{array}{ll}
0 &  \frac{- m^2}{\lambda}  {\rm e}^{-i \Phi} \\
\frac{m^2}{\lambda}  {\rm e}^{i \Phi} & 0
\end{array} \right) \ , &
\partial_- g =  g \left( \begin{array}{ll}
\frac{i}{2}  \partial_- \Phi  & - \lambda \\
\lambda   & \frac{-i}{2}  \partial_- \Phi
\end{array} \right) \ \ .
\end{array}
$$

\section{Heisenberg Model and Nonlinear Schr\"{o}dinger 
Equation}
In this section, we set $X = \RZ^2$ with the origin 
as the base element,   
$(t, x) $ as the coordinates. 
We are going to establish the equivalent relation between  
Heisenberg model and 
nonlinear Schr$\ddot{\rm o}$dinger equation:
$$
\begin{array}{lll}
({\rm HM}) & :  \ \ \partial_0 \vec{S} = 
\vec{S} \times \partial^2_1 \vec{S} \ , & \vec{S} \in 
{\cal C}^\infty(\RZ^2, S^2) \ , \\
({\rm NLS}) &: \ \ i \partial_0 Q + \partial_1^2 Q + 2 |Q|^2Q = 0 \ ,
& Q \in \Omega^0 (\RZ^2)_{\CZ} \ .
\end{array}
$$ 
For an element $(\vec{S}, \vec{t})$ of  
${\cal C}^\infty(X, SO(3))$, 
let $(V, q)$ be the corresponding pair of 1-forms in Lemma 2. 
By (\req(SU2Eq)), we have  
$$
\partial_1^2 \vec{S} = (\partial_A)_1 q_1 \vec{t}_{-} + (\overline{
(\partial_A)_1 q_1})  
\vec{t}_{+} - 4 | q_1|^2 \vec{S} \ , \ \ \ \ A= -iV \ .
$$
Hence one obtains the following constraint of 
the corresponding $(V, q)$ 
for a solution $\vec{S}$ of the equation (HM):
$$
q_0 = i (\partial_A)_1 q_1 \  \ .
$$
Besides the local $H$-gauge symmetries, 
the above condition is invariant under the 
following transformations:
$$
q \mapsto {\rm e}^{2i  v x} q \ , \ \ \ \  
( t , x ) \mapsto \varphi_v (t, x) := ( t , x - 2 v t ) \ ,
$$
here $\varphi_v$ is the Galileo  
tranformation of $\RZ^2$ with $v \in \RZ$. 
Note that we have the relation:
$$
(\varphi_v)_*\left( \begin{array}{c}
 \partial_0  \\
 \partial_1
\end{array} \right) = \left( \begin{array}{c}
 \partial_0 - 2v \partial_1 \\
 \partial_1
\end{array} \right)  \ .
$$
One can eliminate $q_0$ in (\req(SU2FR2)),   
which is reduced to a system of functions $q_1, V_0, V_1$: 
$$
\left\{ \begin{array}{ll}
\partial_0 A_1 - \partial_1 A_0 = 
2 i \partial_1 |q_1|^2 \ ,  &  \\
 i (\partial_A)_0 q_1  + (\partial_A)_1^2  q_1 = 0  \ ,
 &{\rm for} \ A = -iV \ .
\end{array} \right.
$$
By the change of variables, 
$$
W_0 = V_0 - 2 |q_1|^2 \ ,  \ \  W_1 = V_1 \ ,
$$
the above system is equivalent to  
\bea(l)
\left\{ \begin{array}{ll}
\partial_0 B_1 - \partial_1 B_0 = 0 \ , \\
i (\partial_B)_0 q_1 + (\partial_B)_1^2 q_1 + 2 |q_1|^2 q_1  = 0 \ ,  
 &{\rm for} \ B = -iW \ .
\end{array} \right.
\elea(HSF0v)
Under  a gauge transformation $h \in {\cal C}^\infty(X, H)$ 
with $h^{-1}dh = ( d \alpha ) e^3 $, the change of $V_\mu$ in 
(\req(tran)) gives rise the transformations of 
$W_\mu, q_1$, 
$$
q_1  \mapsto  {\rm e}^{i \alpha} q_1 ,  \ \ \ \  
W_\mu  \mapsto  W_\mu +  ( \partial_\mu \alpha ) \ , \ \mu = 0, 1 .
$$
As $W_\mu = \partial_\mu \phi$ for some $\phi 
\in \Omega^0(X)$, a    
solution of (\req(HSF0v)) can always be transfomed into a 
special form of the following type through a local $H$-gauge:
\bea(ll)
(q_1, W_0, W_1) =(Q, 0, 0) \ & ( \Longleftrightarrow 
(q_1, V_0, V_1) = ( Q, 2 |Q|^2 , 0 ) \ ) \ .
\elea(Q00) 
The above function $Q$ satisfies the equation (NLS), and it is
unique up to a multiplicative constant of modulo 1. Hence we have 
obtained the following result:
\par \vspace{0.2in} \noindent
{\bf Theorem 4. } There is an one-to-one correspondence of the 
following data:

(I) $\vec{S}(t, x)$, a solution (HM) with $\vec{S}(0,0)= e$. 

(II) $Q(t, x)$, a solution (NLS), modulo the relation $Q(t, x) \sim c Q(t, x)$ 
with $c \in \CZ , |c|=1$. 
\par \noindent
The above functions $\vec{S}, Q $, are related by 
$$
\left\{ \begin{array}{ll}
\partial_1 \vec{S} & = Q \vec{t}_{-} + \overline{Q} \vec{t}_{+} \ ,  \\
\partial_1 \vec{t}_+ & =-2 Q \vec{S} \ ,  \\
\partial_0 \vec{S} & =  i ( \partial_1 Q) \vec{t}_{-} + 
\overline{i( \partial_1 Q)} \vec{t}_{+} \ , \\
\partial_0 \vec{t}_+  & = -2 i ( \partial_1 Q) \vec{S}
+  2 i |Q|^2 \vec{t}_+  \ ,
\end{array} \right.
$$
for some lifting pair 
$(\vec{S}, \vec{t}) \in {\cal C}^\infty(X,SO(3))$ of   
$\vec{S}$. ( Note that such a lifting always exists, which  
is  unique up to a rotation on tangent planes of 
$S^2$ by a constant angle). 
$\Box$ \par \vspace{.2in} \noindent
If the functions $W_0, W_1$ in (\req(HSF0v)) are  
constant functions, say $
W_0 = \rho \ , \ W_1 = v $, the equation of $q_1$ has the form 
\bea(l)
i ( \partial_0 - 2 v 
\partial_1 )  q_1 + \partial_1^2 q_1 + 2 |q_1|^2q_1 + 
(\rho - v^2 )q_1 = 0 \ .
\elea(NLSv)
Using the $H$-gauge tranformation by setting  
$W +  d (\rho t + v x ) = 0 $, one obtains a solution of (NLS), 
$$
Q ( t, x ) = {\rm e}^{-i ( \rho t + v x)} q_1(t, x) \ .
$$
In the case of $\rho = v^2 $ , one can reduce Eq.(\req(NLSv)) to (NLS) 
by the Galileo  
tranformation $\varphi_v$. 
Hence by starting from one solution $Q$ of (NLS), one produces another 
solution by the relation:
$$
Q_v(t, x) : = (\varphi_v)^*({\rm e}^{i ( v^2 t + v x)}Q) 
 = {\rm e}^{i ( - v^2 t + v x)}
Q ( t, x- 2 v t ) \ .
$$
Therefore we have shown the following result:
 \par \vspace{0.2in} \noindent
{\bf Proposition 7. } The map 
$$
Q ( t , x ) \mapsto  Q_v (t, x) : = {\rm e}^{i ( - v^2 t + v x)}
Q ( t, x- 2 v t ) \ , \ \ v \in \RZ \ , 
$$
defines an 1-parameter family of solutions of the equation (NLS). 
$\Box$ \par \vspace{.2in} \noindent 
One can describe the above family through the following symmetries of 
the $(q, A)$-system:  
$$
q \mapsto {\rm e}^{i ( v^2 t + v x)} q \ , \ \ 
A \mapsto A + i (v^2 d t + v d x) \ , \ \ 
\varphi_v : ( t , x ) \mapsto ( t , x - 2 v t ) \ .
$$
The  expression of $Q$ by $Q_v$ is given by
\bea(l)
Q(t, x) = (\varphi_{-v})^*({\rm e}^{i ( v^2 t - v x)}Q_v) \ .
\elea(QQv)
By (\req(Q00)), one can obtain the zero-curvature representation 
of a solution $Q$ of (NLS) through the process of Section 3. 
In fact, the linear system (\req(gJ)) associated to $Q$ has the form:
\bea(ll)
\partial _1 g = g U_0(t, x )  \ , &  
U_0(t, x ) = \left( \begin{array}{ll}
0 & - \bar{Q}  \\
Q & 0
\end{array} \right) \ , \\
\partial_0 g = g V_0(t, x ) , \ \ & 
V_0 (t, x ) = i \left( \begin{array}{ll}
 |Q|^2 &  \partial_1\bar{Q}  \\
 \partial_1 Q & - |Q|^2
\end{array} \right) \ , 
\elea(NLSsys)
here $g \in {\cal C}^\infty(\RZ^2, SU(2))$. The curvature of the 
above connection $J$ is given by  
$$
F (J) = - i 
 \left( \begin{array}{cc}
0 & -i \partial_0 \bar{Q} + \partial_1^2 \bar{Q} + 2 |Q|^2\bar{Q}  \\
i \partial_0 Q + \partial_1^2 Q + 2 |Q|^2Q & 0
\end{array} \right) 
$$
The Zakharov-Shabat spectral representation of $Q$ is 
the expression, 
\bea(lll)
\partial _1 \psi_{\lambda} = \psi_{\lambda} U(t, x ; \lambda) \ ,&
U(t, x ; \lambda) & = \left( \begin{array}{ll}
\frac{i}{2} \lambda & - \bar{Q}  \\
Q & \frac{-i}{2} \lambda
\end{array} \right) = U_0(t, x ) + \lambda U_1(t, x) \ ,\\ 
\partial_0 \psi_{\lambda} = \psi_{\lambda} V(t, x ; \lambda ) \ , &
V(t, x ; \lambda ) &= i \left( \begin{array}{ll}
 |Q|^2 -\frac{1}{2} \lambda^2 &  \partial_1\bar{Q} -i \lambda \bar{Q}  \\
 \partial_1 Q + i Q \lambda & - |Q|^2 + \frac{1}{2} \lambda^2
\end{array} \right) \\
& &= V_0 (t, x ) + \lambda V_1 (t, x ) + 
\lambda^2 V_2 (t, x ) \ , 
\elea(ZS)
where $U_1 =   e^3  , V_1 = - U_0 , V_2 = - U_1$ and $\lambda \in \CZ$, 
( Note that the $Q$ here is equal to $-i \Psi$ in \cite{FT}  ).
One can relate the above representation to the Galileo symmetries 
of (NLS) and derive it form the $Q_v$-family of 
Proposition 7. In fact, there associates a linear system 
(\req(NLSsys)) for each $Q_v$. With the local $H$-gauge transformation, 
$$
h = {\rm e}^{ \alpha e^3} \ , \ \ 
\alpha (t, x) =v^2 t - v x \ ,
$$
the linear system is transformed into the following one: 
$$
\partial_1 g_{v} = g_{v} \left( \begin{array}{ll}
- \frac{i}{2} v & - {\rm e}^{ -i(v^2 t - v x ) } \bar{Q_v}  \\
{\rm e}^{ i (v^2 t - v x )} Q_v & \frac{i}{2} v
\end{array} \right) \ , \ 
\partial_0 g_{v} =  g_{v} \left( \begin{array}{ll}
 i |Q_v|^2 + \frac{i}{2} v^2 &  i {\rm e}^{ - i (v^2 t - v x )} 
\partial_1\bar{Q_v}  \\
 i{\rm e}^{ i(v^2 t - v x )} \partial_1 Q_v & - i |Q_v|^2 - 
\frac{i}{2} v^2
\end{array} \right) \ . 
$$
Then apply the Galileo transformation $\varphi_{-v}$ to the above system. 
By the relation (\req(QQv)), 
we obtain the expression (\req(ZS)) for $\lambda = -v$ with 
$\psi_{\lambda} = \varphi_{\lambda}^*(g_{-\lambda})$.

\section{Chern-Simons Theory and B\"{a}cklund Transformation}
We are going to discusss the nonlinear 
Schr\"{o}dinger equation of the previous section 
from the 3-dimensional Chern-Simons theory point of view.  
The spectral parameter will be interpreted as a 3 dimensional zero-curvature 
condition through the dimensional reduction. We set 
$X = \RZ^3$ in the discussion of this section with the coordinates 
$(x^0, x^1, x^2)$. Consider the equation
\bea(l)
\partial_0 \vec{S} = \vec{S} \times (\partial_1^2 + \partial_2^2) 
\vec{S} \ , \ \ \ \vec{S} \in {\cal C}^\infty(X , S^2) \ .
\elea(3dS) 
Let $(A, q)$ be the solution of (\req(SU2F)) corresponding to 
$\vec{S}$ in Proposition 3. 
Introduce 
the complex coordinate of $(x^1, x^2)$-plane,
$$
z = x^1 + i x^2 \ .
$$
An 1-form $q$ of $X$ is decomposed into 
$dx^0, dz, d\bar{z}$-components: 
$$
q = q^0 + q' + q'' \ , \ \ \   
q^0 = q_0 dx^0 \ ,  \ q' = q_z dz \ , \
\ q'' =  q_{\bar{z}}d \bar{z} \ . 
$$
Hence one has the corresponding decomposition of the differential $d$ on 
forms of $X$,
$$
d = d^0 + d' + d'' \ .
$$ 
In particular, we have 
the decomposition of a $U(1)$-connection $A$ and the covariant 
differential $d_A$:
$$
\begin{array}{ll}
A = A^0 + A' + A'' & {\rm with } \ \ \overline{A^0} = -  A^0 \ , \ \ 
\overline{A'} = - A'' \ , \\ 
d_A = d_A^0 + d_A' + d_A'' \ ,& 
d_A^0 = d^0 + A^0 \ , \ \ d_A' = d' + A' \ , \ \ 
d_A'' = d'' + A'' \ .
\end{array}
$$
For a solution $(A, q)$ of Eq.(\req(SU2F)),  
its   $(z \bar{z}), (x^0 z ), (x^0 \bar{z})$-components are the 
expressions:  
\bea(ll)
\left\{ \begin{array}{l}
d'A'' + d''A' = 2 ( q'  \overline{q'} + q''\overline{q''}) \ ,  \\
d'_Aq'' + d_A''q' = 0  \ ,
\end{array} \right. \ & {\rm i.e.} \ \ 
\left\{ \begin{array}{l}
\partial_z A_{\bar{z}} - \partial_{\bar{z}} 
A_z = 2 ( |q_z |^2 - |q_{\bar{z}}|^2)  \ , \\
(\partial_A)_z q_{\bar{z}} =  (\partial_A)_{\bar{z}}q_z   \ ,
\end{array} \right. \\
\left\{ \begin{array}{l}
d^0A' + d'A^0 = 2 ( q^0  \overline{q''} + q'\overline{q^0}) \ ,  \\
d^0_Aq' + d_A'q^0 = 0  \ ,
\end{array} \right. \  & 
{\rm i.e.} \ \ 
\left\{ \begin{array}{l}
\partial_0 A_z -  \partial_z A_0 = 
2 ( q_0  \overline{q_{\bar{z}}} - \overline{q_0}q_z ) \ ,  \\
(\partial_A)_0 q_z = (\partial_A)_z q_0 \  ,
\end{array} \right.
\\
\left\{ \begin{array}{l}
d^0A'' + d''A^0 = 2 ( q^0  \overline{q'} + q''\overline{q^0}) \ ,  \\
d^0_Aq'' + d_A''q^0 = 0  \ ,
\end{array} \right. \  & 
{\rm i.e.} \ \ 
\left\{ \begin{array}{l}
\partial_0 A_{\bar{z}} -  \partial_{\bar{z}} A_0 = 
2 ( q_0  \overline{q_z} - \overline{q_0}q_{\bar{z}} ) \ ,  \\
(\partial_A)_0 q_{\bar{z}} = (\partial_A)_{\bar{z}} q_0 \ . 
\end{array} \right.
\elea(CS123)
The corresponding constraint of $(A, q)$ to a solution of (\req(3dS)) is 
given by
$$
q_0 = i 
(\partial_A)_1 q_1 + 
i (\partial_A)_2 q_2  = 4 i (\partial_A)_z q_{\bar{z}} = 4i (\partial_A)_{\bar{z}}  q_z 
\ .
$$
By the relation 
$$
4 (\partial_A)_z (\partial_A)_{\bar{z}} = (\partial_A)_1^2 + 
(\partial_A)_2^2 +2 (\partial_zA_{\bar{z}} - \partial_{\bar{z}}A_z) \ , 
$$
and the change of variables,  
$$
B_0 = A_0 + 4i ( |q_z |^2 + |q_{\bar{z}}|^2) \ , \ \ 
B_1 = A_1 , \ B_2 = A_2 \ , 
$$
Eq. (\req(CS123)) is transformed into the system:   
\bea(l)
\left\{ 
\begin{array}{ll} 
q_0 = 4 i (\partial_B)_z q_{\bar{z}} = 4i (\partial_B)_{\bar{z}}  q_z \ ,
&  \\
\partial_z B_{\bar{z}} - \partial_{\bar{z}} 
B_z = 2 ( |q_z |^2 - |q_{\bar{z}}|^2) \ ,  & \\
\partial_0 B_z -  \partial_z B_0 = 
8 i( \overline{q_{\bar{z}}}  (\partial_B)_z q_{\bar{z}}    + 
q_z \overline{ (\partial_B)_{\bar{z}} q_z } ) 
 - 4 i \partial_z   ( |q_z |^2 + |q_{\bar{z}}|^2) \ ,  & 
z \leftrightarrow \bar{z} \ , \\
i  (\partial_B)_0   q_z
 + ((\partial_B)_1^2 + 
(\partial_B)_2^2 ) q_z + 8  |q_z |^2 q_z = 0 \ , & 
z \leftrightarrow \bar{z} \ .
\end{array} \right.
\elea(CSB123)
Following arguments in \cite{P}, we perform the 
dimensional reduction, and look  
for solutions independent of $x_2$. Denote 
$$
\Psi_+  = \frac{1}{2} q_{\bar{z}} \ , \ \ \Psi_- = 
\frac{1}{2} q_z \ .
$$ 
By the change of variables, 
$$
C_0 = B_0 - i B_2^2 \ , \ \ C_1 = B_1 \ , \ \ C_2 = B_2 \ ,
$$
the system (\req(CSB123)) becomes
\bea(l)
\left\{ 
\begin{array}{l}
\partial_0 C_1 -  \partial_1 C_0 = 0 ,   \\ 
i  (\partial_C)_0   \Psi_\pm
 + (\partial_C)_1^2 \Psi_\pm + 2  |\Psi_\pm |^2 \Psi_\pm = 0 \ ,\\
q_0 = 4 i (\partial_1 +2 C_z )\Psi_+ = 4 i (\partial_1 + 2 C_{\bar{z}}) \Psi_- \ ,
 \\
 \partial_1 C_2 = 16 i (  |\Psi_+|^2 - |\Psi_- |^2 ) \ ,   \\
\partial_0 C_2 = 
-16 [ 
( \overline{\Psi_+} (\partial_C)_1 \Psi_+ - \Psi_+ 
\overline{(\partial_C)_1 \Psi_+} ) 
- ( \overline{\Psi_-} (\partial_C)_1 \Psi_- - \Psi_- 
\overline{(\partial_C)_1 \Psi_-} ) ] \ . \\
\end{array} \right.
\elea(CSC123)
By a suitable $U(1)$-gauge trasnformation, i.e. a  
real-value function $\alpha = \alpha(t,x)$ with
$$
C_0 -i \partial_0 \alpha = 0 \ , \ \ C_1 -i  
\partial_1 \alpha = 0 \ , \ \ {\rm e}^{i\alpha} \Psi_\pm = Q_\pm \ ,
$$
one reduces (\req(CSC123)) into the special form with    
$C_0 = C_1 = 0 $, 
\bea(l)
\left\{ 
\begin{array}{l}
i  \partial_0   Q_\pm
 + \partial_1^2 Q_\pm + 2  |Q_\pm |^2 Q_\pm = 0 \ ,\\
  \partial_1 (Q_+ - Q_- ) 
= i C_2 ( Q_+ +  Q_- ) \ ,
 \\
 \partial_1 C_2 = 16 i (  |Q_+|^2 - |Q_- |^2 ) \ ,   \\
\partial_0 C_2 = 
-16 [ 
( \overline{Q_+} \partial_1 Q_+ - Q_+ 
\overline{\partial_1 Q_+} ) 
- ( \overline{Q_-} \partial_1 Q_- - Q_- 
\overline{\partial_1 Q_-} ) ] \ , \\
q_0 = 4 i (\partial_1 - i C_2 )Q_+ \ \ .
\end{array} \right.
\elea(CSBL)
Solve the above $C_2$ by the expression: 
$$ 
iC_2 = \pm 4  \sqrt{\eta^2 - | Q_+ - Q_-|^2 }  \ ,
$$
where $\eta$ is a constant depending on $Q_+$ and $Q_-$. 
Then the system (\req(CSBL)) is equivalent to 
the B$\ddot{\rm a}$cklund transformation  
of (NLS) for $Q_\pm$:
\bea(l)
\left\{ 
\begin{array}{l}
i  \partial_0   Q_\pm
 + \partial_1^2 Q_\pm + 2  |Q_\pm |^2 Q_\pm = 0 \ ,\\
  \partial_1 Q_+ - \partial_1 Q_-  
=  \pm 4 \eta ( Q_+ +  Q_- ) \sqrt{1 - | (Q_+ - Q_-)/\eta|^2 } \ .
\end{array} \right.
\elea(BL)
Following the above arguments, one can derive a linear system  
on $X$ with the consistency condition (\req(BL)). In fact, 
the three components of the connection $J$ for the 
corresponding linear system (\req(gJ)) are the  
expressions:
$$
\begin{array}{l}
J_0 = 4 i
\left( \begin{array}{cc}
2 \eta^2 + 2 (   Q_+ \overline{Q_-} + 
Q_-\overline{Q_+}) &  
\partial_1 \overline{Q_+} + 4 \eta a \overline{Q_+}  
  \\
 \partial_1 Q_+  +  4 \eta a Q_+   & 
- 2 \eta^2   - 2 (   Q_+ \overline{Q_-} + 
Q_-\overline{Q_+})
\end{array} \right) \ , \\
J_1 = 
\left( \begin{array}{cc}
0 &  
-2 \overline{(Q_+ + Q_-) }  \\
2(Q_+ + Q_-)  & 
0
\end{array} \right) \ , \\ 
J_2 = - 2  \eta i 
 \left( \begin{array}{ll}
 a &   \bar{b}  \\
 b & - a
\end{array} \right) \ , \ 
\end{array}
$$
where $a = \mp  \sqrt{1 -  | (Q_+ - Q_-)/\eta|^2 }$, $  
b = (Q_+ - Q_-)/\eta $.
A key point of this approach is on the relation of  $J_2$ 
and the global gauge tranformations for    
Zakharov-Shabat spectral representations (\req(ZS)) of $Q_\pm$:
$$
\left\{ 
\begin{array}{l}
\partial_1 \psi_+ = \psi_+ U_+(x^0, x^1 ; \lambda) , \\
\partial_0 \psi_+ = \psi_+ V_+(x^0, x^1 ; \lambda )  \ ,
\end{array} \right. \ \ \ \ \
\left\{ 
\begin{array}{l}
\partial_1 \psi_- = \psi_- U_-(x^0, x^1 ; \lambda)  , \\
\partial_0 \psi_- = \psi_- V_-(x^0, x^1 ; \lambda ) \ .
\end{array} \right.
$$
The B$\ddot{\rm a}$cklund transformation is described by 
the existence of a $SU(2)$-valued function $g(x^0, x^1 ; \lambda)$ which 
transforms $\psi_-$ to $\psi_+$:
$$
\psi_+ (x^0, x^1; \lambda) = 
\psi_-(x^0, x^1; \lambda) g(x^0, x^1; \lambda) \ . 
$$
An equivalent form is 
$$
g^{-1} \partial_1 g = U_+ - g^{-1} U_- g \ , \ \ \ 
g^{-1} \partial_0 g = V_+ - g^{-1} V_- g \ .
$$
By the following identification of the spectral parameter $\lambda$ 
with the coordinate $x^2$, 
$$
\frac{4\eta }{\lambda} = \tan  x^2  \ ,
$$
one then obtains the expression of $g(x^0, x^1 ; \lambda)$,  
$$ 
g(x^0, x^1 ; \lambda) = {\rm e}^{- x^2 \frac{J_2}{2\eta}} = 
\left( \begin{array}{ll}
 1 &  0 \\
 0 & 1
\end{array} \right)  \cos  x^2  + 
i \left( \begin{array}{ll}
 a &   \bar{b}  \\
 b & - a
\end{array} \right) \sin  x^2 .
$$

\section{Flat SU(1,1)-Connection and Non-compact $\sigma$-Models }
In this section we discuss models related to the  
$SU(1,1)$-connections. A notable one is the example of E. Witten 
 on the cylindrical symmetry solutions of the 4-dimensional self-dual 
Yang-Mills equation  
in \cite{W}. We are going to follow the same arguments  
of previous sections by replacing the group 
$SU(2)$ by $SU(1,1)$.   
An orthogonal basis of the Lie algebra $su(1,1)$ ( w.r.t. 
the negative Killing bilinear form with signature $(-,-, +)$ ) consists of 
the elements, 
$$ 
e^1 = \frac{-1}{2} \sigma^1 , \ \
e^2 =  \frac{-1}{2}\sigma^2 , \ \
e^3 = \frac{i}{2} \sigma^3  .
$$
The sequence (\req(GAd)) becomes   
$$
1 \longrightarrow \{ {\pm 1} \} \longrightarrow SU(1,1)
\longrightarrow SO(1, 2) \longrightarrow 1 \ .
$$
Consider the diagonal subgroup $H$ of $SU(1,1)$, and identify the 
homogenous space $SU(1,1)/H$ with the hyperboloid in $su(1,1)$ 
having the base element $e= e^3$,
$$
S = \{ \sum_{j=1}^3 x_je^j \ | \ - x_1^2 - x_2^2 + x_3^2 = 1 \} \ . 
$$ 
The tangent bundle ${\bf T}_S$ of $S$ is  a Hermitian line bundle 
with the metric induced by the Killing form of $su(1,1)$. 
Via the stereographic projection, 
$$
p : \BZ^1 \longrightarrow S \ , \ \ \ \zeta \mapsto \vec{S}  \ , \ 
\ \ {\rm with } \ 
\ \ S_1 + i S_2 = \frac{2 \zeta}{1-|\zeta|^2} \ , \ \  \ 
S_3 = \frac{1+|\zeta|^2}{1-|\zeta|^2} \ ,
$$
the hyperboloid $S$ is isometric to the unit disc $\BZ^1$ with the 
Poincar\'{e} metric
$$
 \frac{d \zeta  \otimes d \overline{\zeta}}{ 
(1-|\zeta|^2)^2} \ \ \ .
$$ 
An equivalent model is the upper-half plane,  
$$
\HZ^1 = \{ \tau = x + i y \in \CZ \ | \ y > 0 \} \ , \ \ \ 
ds^2 = \frac{ d \tau  \otimes d \overline{\tau}}{4y^2} \ .
$$ 
Let $J$ be a $SU(1,1)$-connection on a simply-connected manifold $X$,
$$ 
J =   
 - 2 Re(q)e^1 - 2  Im (q)e^2 +  V e^3  =  
q \sigma^- + \overline{q}\sigma^+ + V e^3  \ , \ \ 
q \in \Omega^1(X)_{\CZ} \ , \ \ V \in \Omega^1(X) \ .
$$ 
The zero curvature condition of $J$ can be formulated by the expression: 
\bea(l)
\left\{ \begin{array}{ll}
F(A) = - 2 q  \overline{q}  \ \ ,    & \\
d_A q = 0 \ , \ &{\rm for } \ \ A = - i V \ ,
\end{array} \right.
\elea(SU11F)
which, by (\req(Adeq)), is the consistency condition of the   
linear system of $\vec{S}, \vec{t}$:
\bea(l)
\left\{ \begin{array}{lll}
d \vec{S} & = q \vec{t}_{-} + \overline{q} \vec{t}_{+} \ ,  \\
d_A \vec{t}_+ & = 2 q \vec{S}  \ .
\end{array} \right.
\elea(SU11Eq) 
With the same arguments as in Theorems 1 and 2, one has the 
following results: 
\par \vspace{.2in} \noindent
{\bf Theorem 5.} Let ${\bf L}$ be a Hermitian complex  
bundle over a simply-connected marked manifold $X$. There is an one-to-one correspondence of the 
following data:

(I) $\vec{S} \in {\cal C}^\infty (X, S)$ with $\vec{S}^*({\bf T}_{S}) 
= {\bf L}$ .

(II) $(A, s), $ where $A$ is a $U(1)$-connection on ${\bf L}$, 
$s \in \Omega^1(X , {\bf L})$, satisfying the relation 
\bea(l)
\left\{ \begin{array}{l}
F(A) =  - 2 s s^{\dag}  ,    \\
d_A (s) = 0 . 
\end{array} \right.
\elea(SU11Fi) 
$\Box$ \par \vspace{.2in} \noindent 
{\bf Corollary. } For $X = \HZ$ with the coordinate $z = t + i  r $, 
there is an one-to-one correspondence between the 
following data:

(I) $\vec{S} : X \longrightarrow S $ with $\vec{S}(0,0)= e $, a solution 
of the equation 
$$
\partial_0 \vec{S} = - \vec{S} \times \partial_1 \vec{S} \ .
$$ 

(II) $(A, \Phi)$, where $A$ is a 
$U(1)$-connection, $\Phi \in \Omega^{1,0}(X)$, satisfying the relation:
\bea(l)
\left\{ \begin{array}{l}
F(A) = - 2 \Phi \Phi^*  \ , \\  
d''_A \Phi = 0   \ .  
\end{array} \right.
\elea(SU11LE)

(III) $\phi : X \longrightarrow \RZ \cup \{ - \infty \} $,  a solution of 
Liouville equation:
\bea(l)
 (\partial_0^2 + \partial_1^2)  \phi  = 
8 {\rm e}^{ \phi} \  .
\elea(LEh)  
$\Box$ \par \vspace{.2in} \noindent  
One can derive the expression of 
a general solution of (\req(LEh)) : 
$$
\phi ( z ) =  \log \frac{| \partial_z \zeta (z)|^2 }{(1 - | \zeta (z) |^2)^2} \ \ 
, \ \ \ \zeta: \HZ \longrightarrow 
\BZ \ \ {\rm holomorphic}  \ , 
$$
or equivalently,
$$
\phi ( z ) =  \log \frac{| \partial_z \tau (z)|^2 }{
4 Im (\tau (z))^2} \ \ 
, \ \ \ \tau : \HZ \longrightarrow 
\HZ \ \ {\rm holomorphic}  \ . 
$$
We are going to relate the equations of the above Corollary with the 
self-dual Yang-Mills equation. 
By the change of variables,
$$
A_0 = B_0 + \frac{ i \kappa }{r} \ , \ A_1 = B_1 \ , \ \ 
\Phi = \frac{ \phi }{r} dz  \ , 
$$
Eq.(\req(SU11LE)) is transformed into
\bea(l)
\left\{ \begin{array}{l}
i r^2 (\partial_0 B_1 - \partial_1  B_0 ) =  \kappa   
- 4  \phi \overline{\phi}  \  \\  
 ( \partial_{\bar{z}} + B_{\bar{z}} ) \phi + 
\frac{i(\kappa - 1) }{2r} \phi = 0    \ ,  
\end{array} \right.
\elea(Wsd)
where $\kappa$ is a real constant. One may apply 
a gauge transformation $h \in {\cal C}^\infty(X, H)$ 
with $h^{-1}dh = ( d \alpha ) e^3 $ to the system 
(\req(SU11LE)). It gives rise the  
transformations of functions $B_0, B_1, \phi$, in (\req(Wsd)):
$$
B_\mu \mapsto B_\mu -i \partial_\mu \alpha \ \ \ \  \ ( \mu = 0, 1) \ , 
\ \ \ \phi \mapsto {\rm e}^{i\alpha} \phi \ . 
$$
By setting  
$$
\phi = {\rm e}^{\frac{\psi}{2}} \ \ \ \ {\rm with} \ \ \psi : \HZ \longrightarrow \RZ \ ,
$$
one obtains the expression of $B$,
$$
B = ( \frac{\partial_z\psi}{2}    - 
\frac{i(\kappa - 1) }{2r}) dz - ( \frac{ \partial_{\bar{z}} \psi}{2}
 + \frac{i(\kappa - 1) }{2r}) d\bar{z} \ .
$$
Then $\psi$ satisfies the following Liouville-like equation on a  
curved space-time for all $\kappa$ :
$$
  r^2    (\partial_0^2 +  
\partial_1^2) \psi     
 = - 2 + 8 {\rm e}^{\psi} \ .
$$
With the identification,  
$$
 i B_x = W_x \ , \ 
 i B_y = W_y , \ \ \ 2 \phi = \varphi_1 + i \varphi_2 \ , 
$$
Eq.(\req(Wsd)) for $\kappa = 1 $ is now equivalent to the system,  
$$
\left\{ \begin{array}{l}
 r^2 (\partial_0 W_1 - \partial_1  W_0 ) =  1    
- \varphi_1^2 - \varphi_2^2  \ , \\  
  \partial_0 \varphi_1 + W_0 \varphi_2 = 
\partial_1 \varphi_2 - W_1 \varphi_1 \ , 
\\ 
\partial_1 \varphi_1 + W_1 \varphi_2 = 
-( \partial_0 \varphi_2 - W_0 \varphi_1)  \ ,   
\end{array} \right.
$$
which is the cylindrical symmetric self-dual Yang-Mills equation  
in \cite{W}.

\section{Conclusions} 
We have treated in the present paper only models related to the 
gauge theory of $SU(2)$ or  
$SU(1,1)$, though the formulizm is valid for a 
general gauge group. 
The equivalent relation, established in this work, from 
one site is the well known gauge equivalence between integrable 
models on an arbitary background manifold in the framework of 
Lax pair, while from another site describes 
the $\sigma$-model representations. 
An important point is that there exists a systematic way of producing these 
Lax forms from symmetries of the equation. 
It suggests a possibile  
unified BF gauge theoretical approach 
to an integrable system and its underlying spectral problem 
\cite{BT} \cite{D} \cite{LP} \cite{MPS96} 
\cite{MPS97}. Namely, any 
$\sigma$-model ( on a curved background ) is equivalent to a BF gauge 
field theory in a special gauge. If the model is integrable, then, the 
theory provides a complete description in terms of the Lax pair. Furthermore 
the global gauge structure appears as the spectral parameter, while   
 B\"{a}cklund transformations have the origin in the higher dimensional 
Chern-Simons theory. In this paper we have shown that the hidden 
symmetry of some integrable models with physical interests has its 
structure  
revealed much clearly in a mathematical formulation using the language 
of fiber bundles and connections. 

Another source of main interests in our investigation stems from the 
similarity of Eq.(\req(selfdual)) with the 
Seiberg-Witten equation in the electric-magnetic duality of $N=2$ SUSM 
Yang-Mills theory \cite{SW}. 
Recently, the dimensional 
reduced Seiberg-Witten equations were studied and the set of 
equations shares a similar structure with the Hitchin equation 
except the distinction of Higgs field and Weyl spinor \cite{MR} 
\cite{NS}. In this paper, we have treated the cylindrical symmetric 
self-dual Yang-Mills equation of \cite{W}     
through the framework of integrable systems. 
It is plausible that a similar argument could also apply to  
the equivalent rellation in \cite{FHP} on  the axially 
symmetric Bogomoly monopole equation and Ernst equation 
of general relativity.       
The hope is that our approach based on the exactly soluble models 
could possibly help to illuminate the analytical structure of Seiberg-Witten 
theory. Such a program is now under our investigation. 
\par \vspace{0.4in} \noindent
{\bf Acknowledgements} \par \noindent
We would like to acknowledge helpful discussions with Professor 
J. H. Lee. We would like to thank Academia Sinica at 
Taiwan, and  S. S. Roan would like to thank MSRI and 
Mathematical Department of U. C. Berkeley for the hospitality, where 
this work was carried out.
O. K. Pashaev gratefully acknowledge financial support of 
NSC.

\end{document}